\documentclass[review]{elsarticle}

\usepackage{lineno,hyperref}
\usepackage{booktabs}
\usepackage{amsmath}
\usepackage{amssymb}
\usepackage{multicol}
\usepackage{booktabs,caption}
\usepackage[graphicx]{realboxes}
\usepackage{pdflscape}
\usepackage{natbib}
\biboptions{authoryear}
\modulolinenumbers[0]
 
\journal{Elsevier}

\begin{document}

\begin{frontmatter}

\title{Open source modelling of scenarios for a 100\% renewable energy system in Barbados incorporating shore-to-ship power and electric vehicles}

\author[EUM]{André Harewood}
\ead{andre-gr.harewood@studierende.uni-flensburg.de}
\author[EUM]{Franziska Dettner\texorpdfstring{\corref{corresponding}}}
\cortext[corresponding]{Corresponding author}
\ead{franziska.dettner@uni-flensburg.de}
\author[EUM]{Simon Hilpert}
\ead{simon.hilpert@uni-flensburg.de}

\address[EUM]{Department of Energy and Environmental Management, Europa-Universität Flensburg, Auf dem Campus 1, 24943 Flensburg}

\begin{abstract}
The high dependence on imported fuels and the potential for both climate change mitigation and economic diversification make Barbados' energy system particularly interesting for detailed transformation analysis. An open source energy system model is presented here for the analysis of a future Barbadian energy system. The model was applied in a scenario analysis, using a greenfield approach, to investigate cost-optimal and 100\% renewable energy system configurations. Within the scenarios, the electrification of private passenger vehicles and cruise ships through shore-to-ship power supply was modelled to assess its impact on the energy system and the necessary investment in storage. Results show that for most scenarios of a system in 2030, a renewable energy share of over 80\% is achieved in cost-optimal cases, even with a growing demand. The system's levelised costs of electricity range from 0.17 to 0.36 BBD/kWh in the cost-optimal scenarios and increase only moderately for 100~\% renewable systems. Under the reasonable assumption of decreasing photovoltaic investment costs, system costs of a 100~\% system may be lower than the current costs. The results show that pumped hydro-storage is a no-regret option for the Barbadian power system design. Overall, the results highlight the great potential of renewable energy as well as the technical and economic feasibility of a 100~\% renewable energy system for Barbados.
\end{abstract}

\begin{keyword}
energy transition \sep 100~\% renewable systems \sep SIDS \sep transport electrification \sep open source energy modelling\sep shore-to-ship power
\end{keyword}

\end{frontmatter}

\section{Introduction \& Background}
Energy is key for the well-being and development of all societies. Especially Small Island Developing States (SIDS) are facing numerous social, economic and environmental challenges when it comes to energy. Smaller market size in comparison to larger developed nation counterparts makes diversifying conventional power generation almost impossible, which favours large utility monopolies. In addition, most SIDS lack natural fossil resources and have difficulty diversifying economically \citep{undp_high-level_2018}. Also, SIDS, including Barbados, are playing an increasingly important role pushing for climate action. With a dependence of more than 95~\% on fossil fuel imports, Barbados faces economic vulnerabilities that translate into high electricity prices \citep{henderson_sids_2013}. At the same time, Barbados is the first island in the English-speaking Caribbean to commit to using 100~\% renewable energy \citep{henry_key_2015}. The heavy reliance on fuel imports for energy generation and transportation has affected and is affecting the nation's economic growth and social development.
Barbados has favourable wind and solar resources to aim for a high share of renewable energy sources in the electricity sector as well as the potential to electrify other relevant fossil fuel based sectors. 

Most energy system modelling at the international level and in the SIDS is done with closed black-box energy system models (ESMs). Typically, these are pre-set models and the source code is unavailable for third party review \citep{pfenninger_long-term_2016, hilpert_open_2018}. This makes the analysis of the raw data and the methods used impossible for external analysis, especially for policy planning agencies and researchers. However, under the principles of the Open Energy Modelling (openmod) initiative, the data,code and documentation for the model used in this analysis is shared publicly. Open science should be the standard to promote transparency in scientific investigations. Increased openness is also significant to foster open and frank dialogue between SID state governments and the lending agencies that often require energy modelling investigations as a prerequisite for policy based loans \citep{atteridge_development_2019}. Whereas only a few studies address the possibility of a 100~\% RES for Barbados, none utilise an open source energy system model to create transparent and reproducible results. This study fills the above mentioned research gap by developing an open energy system model based on the Open Energy Modelling Framework (oemof) \citep{hilpert_open_2018} for Barbados. This allows to determine, which share of renewable energies is technically and economically feasible in the future electricity system of Barbados in 2030, as well as modelling not yet electrified sectors, such as passenger transport and the cruise tourism sector. Applying a greenfield approach to energy system modelling, which removes all boundary conditions by today's systems to achieve the best overall system performance \citep{geidl_greenfield_2006}, the present analysis is an indispensable prerequisite for future, detailed power system planning. 

For an in depth analysis and a more profound understanding of the Barbadian energy system, including currently installed capacities, demand analysis, renewable energy potentials and political framework conditions please consult the supplementary material. The most relevant information is summarised below. Weather conditions in Barbados are promising for a cost efficient RES. The dynamic of average wind and solar daily capacity factors is shown in Figure \ref{fig:capacity-factor} based on selected weather years.

\begin{figure}[!h]
    \centering
    \includegraphics[width=0.9\textwidth]{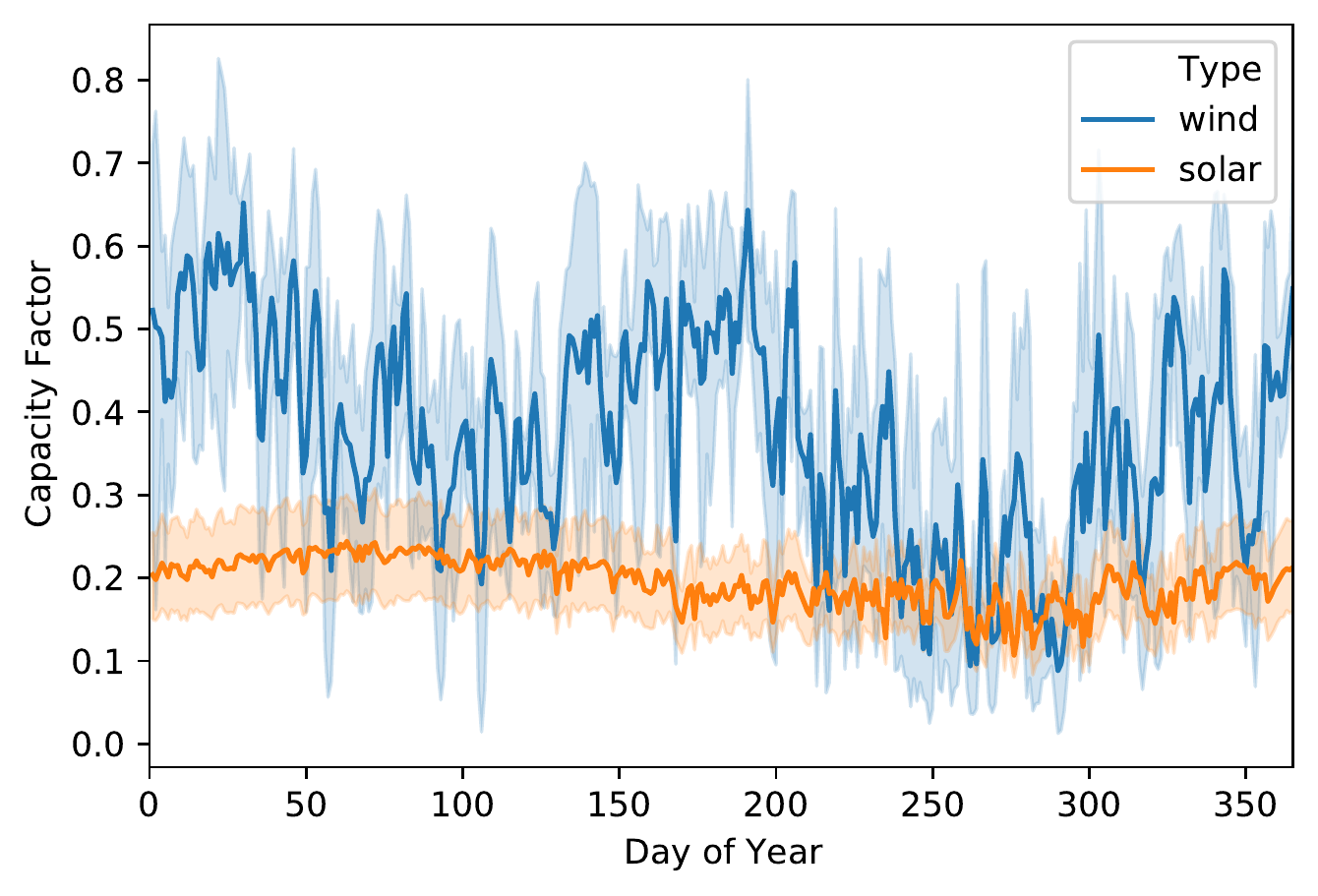}
    \caption{Average wind and solar daily capacity factor and 95~\% confidence interval of four years (2002, 2004, 2010, 2014) based on \emph{Renewables.Ninja} \citep{pfenninger_long-term_2016} data for location at 13.32 latitude and -59.6321 longitude.}
    \label{fig:capacity-factor}
\end{figure}

The relevance of bioenergy for a 100~\% RES in Barbados to complement the wind and solar potential has been noted in \cite{espinasa_achieving_2016, gob_barbados_2019} and \cite{irena_barbados_2016}, among others. Bagasse as a residue from the sugar production process can be used for the generation of bio-energy \cite{marshall_bio-digestion_2019}. As an island system aiming for a 100~\% renewable energy supply, it is natural to include sectors traditionally powered by fossil fuels in the modelling. The passenger transport sector as well as the cruise sector as an important tourism sector with a total contribution of almost 40~\% to the GDP as well as the national employment in 2015 is included in the modelling approach.

\subsection{Research Question}
The Barbadian situation puts a strong emphasis on climate change mitigation because of the small island characteristics and economical challenges connected to the fuel import dependency. The research questions in this paper are connected to a techno-economic assessment of the energy future of 2030 and built on \autoref{state_of_research}. A representative year was selected with perfect foresight to model the system and its costs. Neither the transition nor cost during transition are included, as this is not the goal of the chosen approach. This is a trade-off between complexity and simplicity. For this purpose, an open source model based on the Open Energy Modelling Framework (oemof) was developed, utilising a greenfield modelling approach, aiming to answer the following research questions: 

(1) What is the cost-optimal share of renewable energy sources technically feasible for the Barbadian electricity system?

(2) What are suitable combinations of storage technologies and particularly the role of PHS in a 100~\% RES for Barbados? 

(3) What are the techno-economic effects in the electricity system due to the electrification of the passenger transport sector and shore-to-ship power supply for cruise ships in Barbados?

\section{State of research}
\label{state_of_research}
Several studies have highlighted the impact of fossil fuel dependency on the SID state characteristics that directly and indirectly exacerbate the need for sustainable development. This compromises the ability of Barbados to invest in sustainable development initiatives to reduce the dependency. According to \cite{blechinger_barriers_2014}, Caribbean islands face several barriers for the development of renewable energy technologies, which can be clustered in technical, economic, political and social. 

Scenario analyses have been conducted on possible futures for the Barbadian electricity system solely in studies on behalf of the Government of Barbados. A first investigation was carried out under the Sustainable Energy Framework (SEF 2010) for Barbados \citep{iadb_support_2010}. In 2010, a 100~\% RES was not yet an official policy target for the Government and was therefore not analysed. The study results led to the initial policy target of 29~\% of renewable energy in the electricity mix for Barbados in 2030 \citep{iadb_sustainable_2010}. However, the report did not publish modelling code, tools or methodologies. The Government also consulted the International Renewable Energy Agency (IRENA) for scenario analysis of the power generation sector, which recommended a share of 76~\% renewable energy instead of a 100~\% RES. The study did not consider pumped hydro-storage (PHS) as a viable storage option to support the 100~\% RES, but battery storage technologies \citep{irena_renewable_2019}. The viability of the 100~\% RE scenario depends on utilising a biomass potential of 54~MW, which would require 16 million tonnes of sugarcane per year from 20,000 hectares of land \citep{irena_barbados_2016}. However, as recent as 2016, only 7,000 hectares of arable land for sugarcane production were available \citep{lind_potential_2018}. A study conducted by \cite{hohmeyer_economic_2017} examined the possibility of a 100~\% RES using PHS to achieve a dispatch with lower levelised costs of electricity. All scenarios employed between 200 and 260~MW of wind and PV, as well as 11~MW of solid waste combustion. The scenarios varied only in the extent of biomass utilisation and technology for bio energy generation. Within the Barbados National Energy Policy (BNEP) 2019, scenario analysis was also conducted to examine possible dispatch options. Although, a 100~\% RE system remains the goal for Barbados, the purpose of the scenario analysis was to examine possible dispatch options of a 76~\% RE system, centred on \cite{irena_barbados_2016}, using a multi-criteria approach based on environmental, economic and social considerations \citep{gob_barbados_2019}. \cite{irena_barbados_2016} is the only study to examine an optimised dispatch with renewables, conventional generators and battery storage using LEAP and OSeMOSYS. Table \ref{tab:overview_of_pre_studies} summarises key figures of the previously introduced scenario analyses.

\subsection{Scientific contribution}\label{sci_contribution}
Models of a 100~\% RES for Barbados exist, are, however, based on closed energy system models, which pose challenges associated with the inability to reproduce the outputs for external examination. Previous studies are also heavily depended on external aid, without which investment in most SIDS energy sectors would be non-existent \citep{niles_small_2013}. Donor agencies have the technical, legal and professional capacities to draft policies and review institutional structures. This may also lead to the development of energy policies as a prerequisite or condition to further access funds from the donor or loan agency. When using closed models as seen in the development of the SEF 2010 for Barbados, the findings are solely presented as final results, without scrutiny from third parties.

Open access research may be more beneficial for the Barbadian energy system by improving transparency, generating and fostering the re-usability of results as well as adding scientific value to discussions on carbon neutral energy systems. The greater openness offered by oemof \citep{hilpert_open_2018}, in the form of open source code and raw model results, can address the problem of external donor agencies and the possibility of biased results. As shown in \autoref{state_of_research}, none of the previous studies considered high shares of RE, the possibility of PHS nor a combination of multiple storage types. Additionally, the study at hand is the first to assess the possibility of electrifying one of the most relevant revenue sectors for Barbados, the cruise ship tourism sector. Analysing shore-to-ship renewable power can benefit the country and the society as a whole with lower GHG emissions as well as reduced air pollution levels and increased health standards. The investigation focuses on examining various possible scenarios for a cost-optimal integration of RE into the energy system, as well as analysing the possibility of a 100~\% RES. Furthermore, the presented open source model can be used to address similar research questions for other SIDS energy systems. Therefore, the presented model does not only contribute to the scientific debate of decarbonising energy systems to mitigate climate change, but can also mark an important step for capacity building and sustainable development in other SIDS through open, unbiased dialogue.

\section{Mathematical model}
To assess the future energy system, a bottom-up optimisation linear programming model is applied. The model has been implemented based on the model generator oemof-solph \citep{krien_oemofsolphmodel_2020} using the oemof-tabular interface \citep{hilpert_introducing_2021}, which are both part of the Open Energy Modelling Framework (oemof) \citep{hilpert_open_2018}. A similar model has been applied for the analysis of the Jordanian energy system \citep{hilpert_analysis_2020}.\\

In the following mathematical description, endogenous (input) variables are printed in bold to distinguish them from exogenous model variables. The model minimises total operational costs for the time horizon $T$ and annualised investment costs of units $u \in U$ as well as storage investment costs of all storages $s \in S$ for the Barbadian electricity system. The respective objective function is given below in \autoref{eq:objective}.

\begin{align}
    \text{min} : \overbrace{\sum_{t \in T} \sum_{u \in U}  c^{opex}_u  \textbf{p}_{u,t}}^{\text{operational cost}} \cdot \tau  +  \overbrace{\sum_{u \in U} c^{capex,p}_u  \textbf{p}^{nom}_u}^{\text{power inv. cost}}   + \overbrace{\sum_{s \in S} c^{capex, e}_s  \textbf{e}^{nom}_s}^{\text{energy inv. cost}} \label{eq:objective}
\end{align} 

The operational costs $c^{opex}_u$ are calculated based on the efficiency $\eta_u$ of a unit $u$ and its fuel cost $c^{fuel}_u$ according to \autoref{eq:opex}. 
Annualised investment costs $c^{capex}$ are calculated based on the lifetime $n$, weighted cost of capital $wacc$ and the specific investment cost of a technology $capex_u$ as well as the fixed operation and maintenance cost $fom$ in \autoref{eq:capex}. 

\begin{align}
    c^{opex}_u &= \frac{c^{fuel}_u +c^{var}_u}{\eta_u} \label{eq:opex} \\
    c^{capex}_u &= capex_u \cdot
            \frac{(\text{wacc} \cdot (1+\text{wacc})^n)}
            {((1 + \text{wacc})^n - 1)} \cdot (1 + fom) \label{eq:capex}
\end{align}

The demand must equal the sum of supply of all producing units as described in \autoref{eq:demand-constraint}. Note, that for the storage units, $p$ can also take negative values when the storage is charging. The total demand in every time step is composed of different loads such as households, electrical vehicles and cruise ships with their specific patterns and is assumed to be inelastic.

\begin{align}
     \sum_{u \in U} \textbf{p}_{u, t} \cdot \tau  = \sum_{l \in L} d_{l, t} \cdot \tau +  \textbf{p}^{excess}_t \cdot \tau\qquad \forall t \in T
     \label{eq:demand-constraint}
\end{align}

For all investment units, the supply is limited by the installed nominal power \textbf{p}$_i^{nom}$ described in \autoref{eq:p-bounds}, which is bounded by a lower and upper investment limit as shown in \autoref{eq:power-potential}. 

\begin{align}
    0 \leq \textbf{p}_{u, t} \leq \textbf{p}^{nom}_{u}  &\qquad \forall u \in U, t \in T
    \label{eq:p-bounds}\\
    \underline{p}_i \leq \textbf{p}^{nom}_u \leq \overline{p}_i &\qquad \forall u \in U
      \label{eq:power-potential}
\end{align} 

The energy storage balance in \autoref{eq:storage-balance} is applied for all modelled storage types. The balance includes standing losses $\eta^{loss}$ as well as charge and discharge efficiencies $\eta^{in/out}$.

\begin{align}
    \textbf{e}_{s, t} = \textbf{e}_{s, t-1} \cdot \eta^{loss}_s - \frac{\textbf{p}^{out}_{s,t}}{\eta^{out}_{s}} \cdot \tau   + \textbf{p}^{in}_{s,t} \cdot \eta^{in}_{s} \cdot \tau \qquad \forall s \in S, t \in T
    \label{eq:storage-balance}
\end{align} 

Additionally, the power of the storage is limited by the optimised nominal power shown in \autoref{eq:storage-p-bounds}. 

\begin{align}
       - \textbf{p}^{nom}_s \leq \textbf{p}_{s,t} \leq \textbf{p}^{nom}_{s} &\qquad \forall s \in S, t \in T 
       \label{eq:storage-p-bounds}
\end{align}

For all volatile RE technologies, i.e. PV and wind, the power output is determined by \autoref{eq:re-power} where $c^{profile}_{v,t}$ is the time dependent normalised generation profile of the unit $v \in V$.

\begin{align}
       \textbf{p}_{v,t} = c^{profile}_{v,t} \textbf{p}^{nom}_v &\qquad \forall v \in V, t \in T 
       \label{eq:re-power} 
\end{align}

Analogous to \autoref{eq:p-bounds} and \ref{eq:power-potential}, the energy storage content and its maximum investment is bounded as shown in \autoref{eq:storage-e-bounds} and \ref{eq:energy-potential}. 

\begin{align}
   e^{min}_{s} \cdot \textbf{e}^{nom}_{s}\leq  \textbf{e}_{s,t} \leq \textbf{e}^{nom}_{s} &\qquad \forall s \in S, t \in T
   \label{eq:storage-e-bounds} \\ 
   0 \leq \textbf{e}^{nom}_{s} \leq \overline{e}_s &\qquad \forall s \in S 
   \label{eq:energy-potential}
\end{align}  

The dispatchable renewable units $d \in D$ are modelled with a conversion process as described in Equation \ref{eq:conversion}. 

\begin{align}
       \textbf{p}_{d,t} = \eta_{d} \cdot \textbf{h}_{d,t}  \qquad \forall d \in D, \forall t \in T
       \label{eq:conversion}
\end{align}

The conversion process allows to introduce the input of fuel $h$, which can then be bounded for a time horizon within \autoref{eq:fuel-max}. This equation allows to model (annual) resource limitations in biomass or waste.

\begin{align}
       \sum_{t \in T} \textbf{h}_{d,t} \cdot \tau \leq \overline{h}_c \qquad \forall d \in D
       \label{eq:fuel-max}
\end{align}

To model RE penetration within the system by an exogenously defined RE share, an additional constraint is introduced. The renewable energy share is defined within \autoref{eq:RE-share} by the share of conventional technologies $c \in C$.

\begin{align}
    \sum_{t \in T} \sum_{c \in C}  \textbf{p}_{c,t} \cdot \tau \leq (1-RE^{share}) \cdot  \sum_{l \in L} c^{amount}_{l} 
    \label{eq:RE-share}
\end{align}

Finally, the excess supply within the model is limited by two equations. Equation \ref{eq:excess_power} limits the excess power in every time step by to 10~\% of the peak demand $d^{peak}$ of the year, while Equation \ref{eq:excess_energy} limits the excess energy for the whole time horizon. 

\begin{align}
    \textbf{p}^{excess}_t &\leq 0.1 \cdot d^{peak} \qquad \forall t \in T \label{eq:excess_power}\\
    \sum_{t \in T}  \textbf{p}^{excess}_t \cdot \tau &\leq 0.1 \cdot \sum_{l \in L} c^{amount}_{l}
    \label{eq:excess_energy}
\end{align}

\subsection{Scenario assumptions}
The analysis considers ten scenarios, of which the main parameters are summarised in \autoref{tab:study_summary}. All scenarios are modelled in a cost-optimal (without \autoref{eq:RE-share}) and a 100~\% renewable case, where \autoref{eq:RE-share} applies with a value of 1 for the $RE^{share}$ parameter.

A greenfield approach is applied, which is a standard procedure in energy system modelling. Greenfield planning largely neglects the constraints given by today's system and future planning, except for natural limits such as wind and solar resources \citep{geidl_greenfield_2006}. Most of the currently installed power plants will retire in 2030 due to age. The first scenario represents the status-quo (SQ) on the demand side and is primarily designed for the comparability of results with other studies and therefore does not consider electrification of sectors other than energy generation. The second scenario is used as a reference scenario (REF) for the remaining scenarios and includes the electrification of passenger transport vehicles and cruise ships.

The electricity demand in Barbados was modelled based on the assumptions of the Integrated Resource Plan \citep{blp_integrated_2014}. Although, the document has not been updated since 2014 it remains the primary source of information on generation and future demand of the Barbadian energy system. \cite{irena_barbados_2016}, \cite{hohmeyer_economic_2017} and \cite{gob_barbados_2019} were used to validate and compare the final demand in the target year 2030. The annual hourly load profile was simulated from a sample 7-day hourly demand curve from 2014, as outlined within \cite{hohmeyer_economic_2017}. The total future system demand for 2030 was set at 943~GWh, which is the current demand of 2019. The demand increased only marginally from 912~GWh in 2013 and, as stated in \cite{blp_integrated_2014}, is not expected to increase substantially due to demand side management and energy efficiency measures in the residential and commercial sector. The high demand scenario (HD), assumes, analogous to \cite{blp_integrated_2014}, an expected growth in residential, commercial and industrial demands between 2012 to 2030 of 1.2\% annually, which generates an annual system demand of 1,321.3~GWh for 2030.

The demand profiles for the electrification of the transport sector and the cruise ships are analysed as separate profiles to the annual system demand. All scenarios, except SQ and EVUC, use the same demand profiles for the electrification of sectors as well as the same base demand (expect HD) with varying dispatchable renewable generation, PHS and battery storage and costs to address uncertainties. 

On the renewable supply side, wind, solar, biomass (bagasse) and waste are considered in all scenarios with varying capacities. Based on \cite{hohmeyer_economic_2017}, the annual potential of biomass and waste is limited in the REF scenario to 656~GWh\textsubscript{th} and 218~GWh\textsubscript{th} respectively (169~GWh\textsubscript{el} and 74~GWh\textsubscript{el}). The generation profiles for wind and solar are based on the \emph{Renewables.Ninja} project \citep{pfenninger_long-term_2016, staffell_using_2016}. The conventional units considered are low and medium speed diesel generators (lsce and msce) fired with heavy fuel oil (hfo). A combination of fuel efficient lsce and smaller msce is considered as a viable solution for Barbados, according to \citep{blp_integrated_2014}. The model and the scenario input data are publicly available on GitHub \citep{hilpert_oemof-barbados_2021}.

\begin{table}[!h]
\centering
\footnotesize
\caption{Overview of scenario assumptions including demand and technical parameters. All scenarios, except SQ, consist of base demand, electric vehicle demand and cruise demand as in the REF scenario - with variations as stated in parameters.}
\label{tab:study_summary}
\begin{tabular}{p{3cm}p{1cm}p{4.5cm}p{2cm}}
\toprule 
Scenario    & Symbol    & Parameter & Value\\ 
\midrule 
Status-quo              & SQ    & Base demand   & 943~GWh\\
Reference               & REF   & Base demand   & 943~GWh\\
                        &       & EV demand, controlled charging     & 265~GWh\\
                        &       & Cruise demand & 44.15~GWh \\
                        \midrule
High Demand             & HD    & Base demand +\,1.2\%/a & 1,321.3~GWh\\
Restricted Biomass      & RB    & Biomass potential -50\%   & 328~GWh$_{th}$ \\
                        &       & Waste potential -50\%     & 109~GWh$_{th}$  \\
No PHS & NPHS  & No PHS investment                   & - \\ \midrule                 
\multicolumn{4}{c}{\textbf{Cost variations}}\\ \midrule
Low oil price           & LOP   & Oil price -50\%           & 59.2 \$/kWh \\
Electric vehicle uc     & EVUC  & EV demand, uncontrolled charging & 265~GWh \\
Low RE costs    & LRC   & Long term wind costs   & 2900 \$/kW\\
                        &       & Long term PV-distributed costs & 2100 \$/kW \\
                        &       & Long term PV-utility costs     & 1500 \$/kW \\
Medium RE costs    & MRC   & Medium term wind costs   & 3335\$/kW\\
                        &       & Medium term PV-distributed costs & 3150 \$/kW \\
                        &       & Medium term PV-utility costs     & 2250 \$/kW \\      
High bagasse cost       & HBC      & Higher bagasse investment costs  & 18400 \$/kW\\
\bottomrule
\end{tabular}
\end{table}

\subsubsection*{Cruise and Vehicle Demand}
\begin{figure}[!h]
    \centering
    \includegraphics[width=\textwidth]{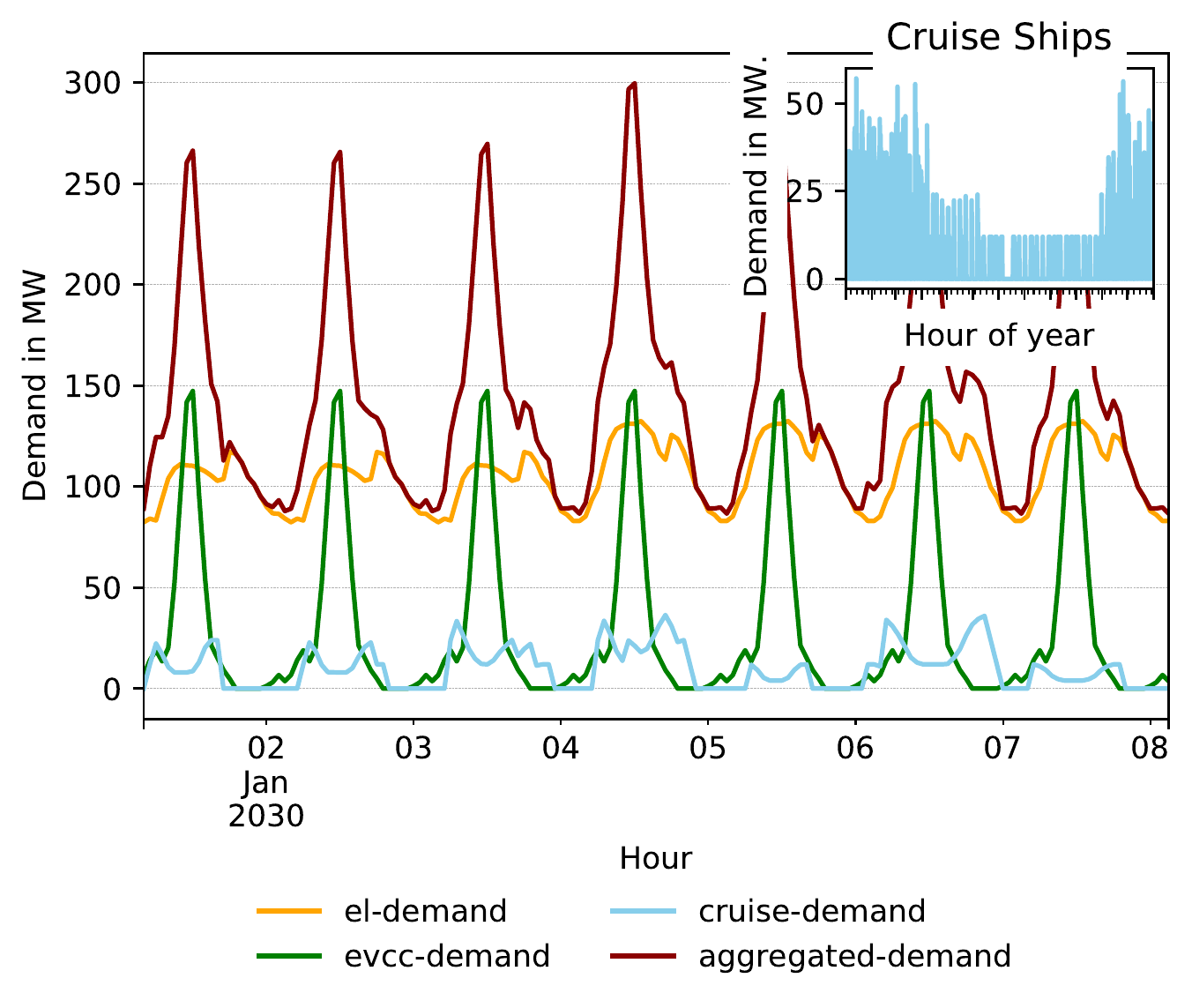}
    \caption{Modelled load patterns of electricity demand, electric vehicles and cruise ships for one week (main figure) and the cruise ship daily average for a whole year (small figure upper right).}
    \label{fig:load-pattern}
\end{figure}

The demand profiles for electric vehicles and shore-to-ship charging for cruise ships were created as separate demand profiles, as depicted in \autoref{fig:load-pattern}. At the time of the present analysis, no information regarding the energy consumption nor the demand profile of cruise ships in Barbados were publicly available. However, \cite{hoyte_shore--ship_2016} has conducted an extensive study of the cruise industry demand in Barbados, which is used in this research as well as information from the local port authority \citep{bpi_port_2020}. Analysis of the port data shows, that cruise ships typically dock for a period of 5 to 20 hours, with 92~\% docking longer than 10 hours. About 50~\% of all recorded cruise arrivals (431 in 2018) docked between 10 and 12 hours, only 7.7~\% stayed less than 10 hours. The large majority of all ships arrived between 5:00 and 10:00 am (see \autoref{fig:arrival_and_stay}). A 12 hour demand profile for the docking time of one generic ship was applied, using a peak demand of 12~MW, multiplied by the actual arrival data from 2018 \cite{bpi_port_2020}. Within the first two hours after docking with peak demand, the demand drops to low demand, increasing again to peak demand at the end of the docking time. A peak demand of 15~MW was assumed for all cruise ship types \cite{hoyte_shore--ship_2016}. The seasonal pattern in the annual demand from cruise ships is clearly visible in \autoref{fig:cruise_demand}. 

On Barbados, there were 94,100 registered vehicles as of 2016 \citep{irena_barbados_2016}. Just over 81~\% are passenger vehicles, followed by light and heavy goods vehicles (11.2~\%), private taxis and minibuses (3~\%) as well as rental cars (2.7~\%). Following \cite{irena_barbados_2016}, only the electrification of the largest share of vehicles, passenger transport, was considered. Within the model, in all scenarios but the SQ scenario, an electrification rate of 80~\% by 2030 was assumed, which is in-keeping with the goals of the BNEP 2019. Controlled charging is mapped in all scenarios but SQ and EVUC, where peak demand occurs around noon. The annual demand of 265.4~GWh as well as the daily load profile was modelled after \cite{irena_barbados_2016} and \cite{gay_small_2018}. As a comparison, a scenario with uncontrolled charging (EVUC) is set up, where the charging peak occurs around 6:00 pm.

The electrification scenarios cause a higher total demand and an altered aggregated demand pattern. Figure \ref{fig:load-pattern} shows the first week of 2030 with three different modelled load patterns as well as the aggregated demand with values of the REF scenario.

\begin{table}[!h]
\centering
\caption{Cost and technical data for supply units in the REF scenario with costs in BBD. wacc - weighted average cost of capital, $\eta$ - efficiency, fom - fixed operation and maintenance, vom - variable operation and maintenance. Values from \cite[p. 76/92]{blp_integrated_2014} for generation technologies and \cite{mongird_energy_2019} for storage units.\\ $^*$Storage cost in include 100 \$/kW and 600 \$/kWh with a E/P ratio of four.\\ Note: A sensitivity analysis for wacc of 4\% and 10\% is included to reflect uncertainties in assumptions.}
\label{tab:tech_data}
\begin{tabular}{llp{1cm}p{0.7cm}p{1cm}p{1cm}p{0.8cm}p{1.2cm}p{1cm}}
\toprule
      &      &  capex (\$/kW) &  life \newline(a) &  wacc (-) &  $\eta$ \newline(\%) &  fom (\%) &  vom (\$/MWh)  & size (MW)\\
Carrier & Technology &               &               &           &          &          &       &       \\
\midrule
wind & onshore &        3500 &            20 &      0.07 &    100.0 &      4.0 &            0 & 1 \\
solar & pv-distributed &        5400 &            20 &      0.07 &    100.0 &      1.0 &            0 & - \\
      & pv-utility &        3900 &            20 &      0.07 &    100.0 &      1.0 &            0 & 1\\
lithium & battery &         2500$^{*}$ &            12 &      0.07 &     94.9 &      3.0 &            0 & - \\
hydro & phs &        5000 &            45 &      0.07 &     89.4 &      1.0 &            0 & - \\
hfo & lsce &        2853 &            30 &      0.07 &     46.4 &      4.0 &           12 & 31.7\\
      & msce &        2344 &            25 &      0.07 &     43.9 &      7.0 &           18 &17.1 \\
bagasse & st &        8000 &            25 &      0.07 &     25.3 &      3.6 &           15 & 25 \\
waste & ocgt &       18000 &            20 &      0.07 &     34.1 &      3.9 &           15  & 13.5 \\
\bottomrule
\end{tabular}
\end{table}

\newpage

\section{Results}

\subsection{Installed capacities and energy balance}
\begin{figure}[!h]
    \centering
    \includegraphics[width=\textwidth]{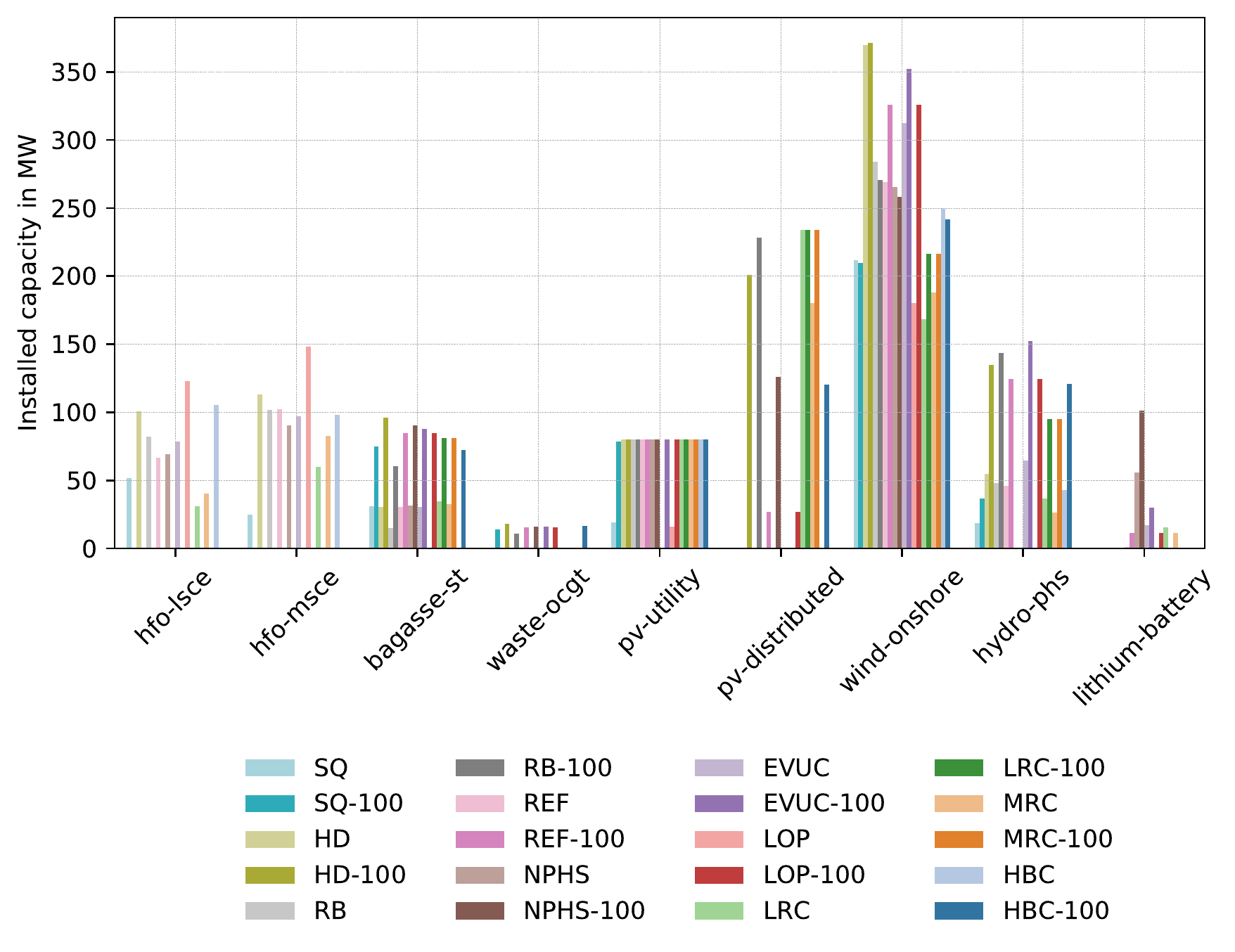}
    \caption{Installed capacities for all scenarios in the cost-optimal and 100\% RE setup.}
    \label{fig:installed_capacities}
\end{figure}

The results of this study have to be critically read in the light of the used method and model. Figure \ref{fig:installed_capacities} depicts the installed capacities per supply technology. In all scenarios, independent of the cost-optimal or the 100~\% renewable energy (RE) case, RE sources are the substantial share of the overall installed capacity. Wind energy has the highest capacities ranging from around 168~MW in the LRC-scenario to 371~MW in the HD-100 scenario. Within the SQ scenario, 211~MW wind capacity compared to 269~MW in the REF scenario are necessary to cover the increased demand due to the electrification of passenger transport and cruise ships. Additionally, much higher conventional capacities are needed in the REF scenario (169~MW msce) compared to the SQ scenario (76~MW), as well as increased PHS (46~MW respective 18~MW). The maximum possible PV-utility potential of 80~MW is exploited in all scenarios, but SQ and SQ-100. Due to its investment cost, PV-distributed reaches high capacities only in the 100~\% scenarios, either due to cost reductions (MRC-100, LRC-100), higher demand (HD-100) or a restricted biomass potential (RB-100). Within the scenario of uncontrolled EV charging, no distributed PV is installed due to the higher costs of integrating PV electricity. 

\begin{figure}[!ht]
    \centering
    \includegraphics[width=\textwidth]{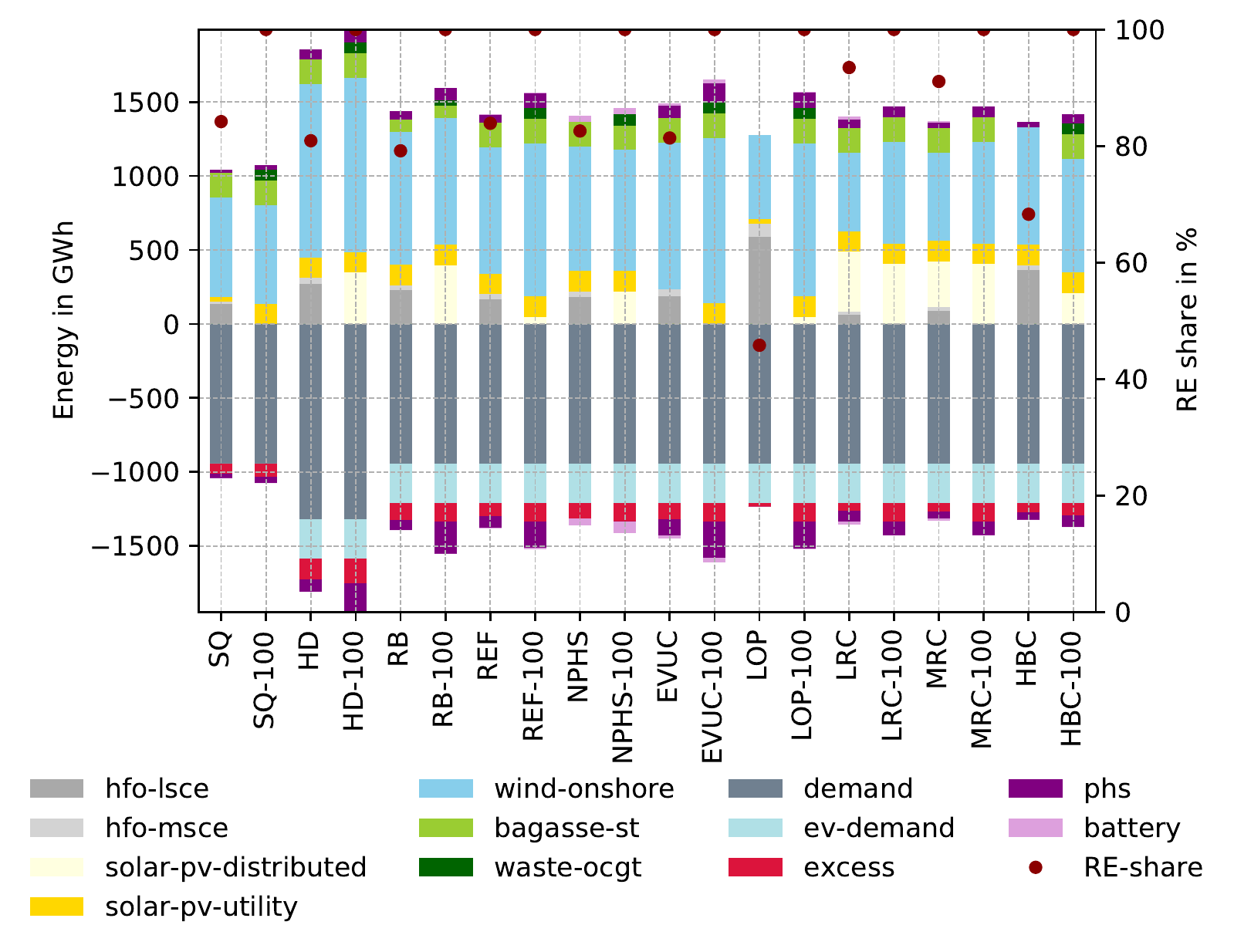}
    \caption{Energy supply and demand balance (left axis) and renewable energy share (right axis) for all scenarios.}
    \label{fig:energy}
\end{figure}

Despite their high investment costs, bagasse capacities around 30~MW can be found in all cost-optimal scenarios, except RB with 15~MW and no investment with a lower oil price (LOP) or higher investment cost of biomass (HBC). The reason are the significantly lower fuel costs of bagasse compared to heavy fuel oil, which has lower investment cost but high operational cost. Hence, for the LOP scenario, no bagasse investment is cost-optimal. In all 100~\% RE scenarios, bagasse capacity increases to values ranging from 60~MW in the RB-100 scenario to 96~MW in the HD-100 scenario. Installed capacities within SQ compared to REF are almost identical, with 30~MW (74~MW respective 84~MW in the 100~\% cases). The lower bagasse investment in the RB-100 scenario due to the restricted biomass potential is compensated by high PHS capacities. In all 100~\% scenarios, PHS capacity is above 95~MW (except \~37~MW in SQ). However, an installed capacity of 143~MW and 152~MW highlights the relevance of the PHS in the case of restricted biomass potential (RB-100) and uncontrolled EV charging (EVUC-100), respectively. Due to the large PHS capacity in the 100~\% scenarios, battery storage capacity is only required for up to 15~MW in cost-optimal cases. Lithium batteries also become important within the EVUC scenarios, to meet the shifted charging demand in the evening and PV peak during the day. Also, within the REF scenarios, batteries are applied (1~MW REF and 11.6~MW REF-100), whereas in the SQ scenarios, no battery capacities are necessary. If there is no PHS investment (NPHS), battery capacities of 56~MW in the cost-optimal and 101~MW need to be included. No conventional capacities exist in the 100~\% scenarios. In the cost-optimal scenarios values of conventional installed capacities are ranging from 76~MW in the SQ scenario to 271~MW in the LOP scenario. Investment in waste units can only be observed for 100~\% RE scenarios due to the high investment costs of 18.000 BBD/kW. Investment in lithium batteries is necessary in the NPHS scenarios, as well as, the EVUC scenarios, where the PV generation pattern cannot compensate for the increased demand from electric vehicles. The corresponding energy balance and RE-share is shown in Figure \ref{fig:energy}. 

The cost-efficient investment of renewable energy leads to high RE-shares above 80~\% in most cost-optimal scenarios (except for the LOP with 46~\% and HBC with 68~\%) with a maximum of around 93~\% in the LRC scenario. With the favourable capacity factor of wind and its therefore large installed capacities, the demand is covered mainly by wind, followed by conventional supply in the cost-optimal scenarios. Only with a reduction in PV cost (LRC, MRC), PV supplies the second largest share of electricity in the cost-optimal cases. Bagasse and waste supply only increase slightly from the cost-optimal to the 100~\% cases. 

\subsection{Dispatchable Operation}
\begin{figure}[!h]
    \centering
    \includegraphics[width=\textwidth]{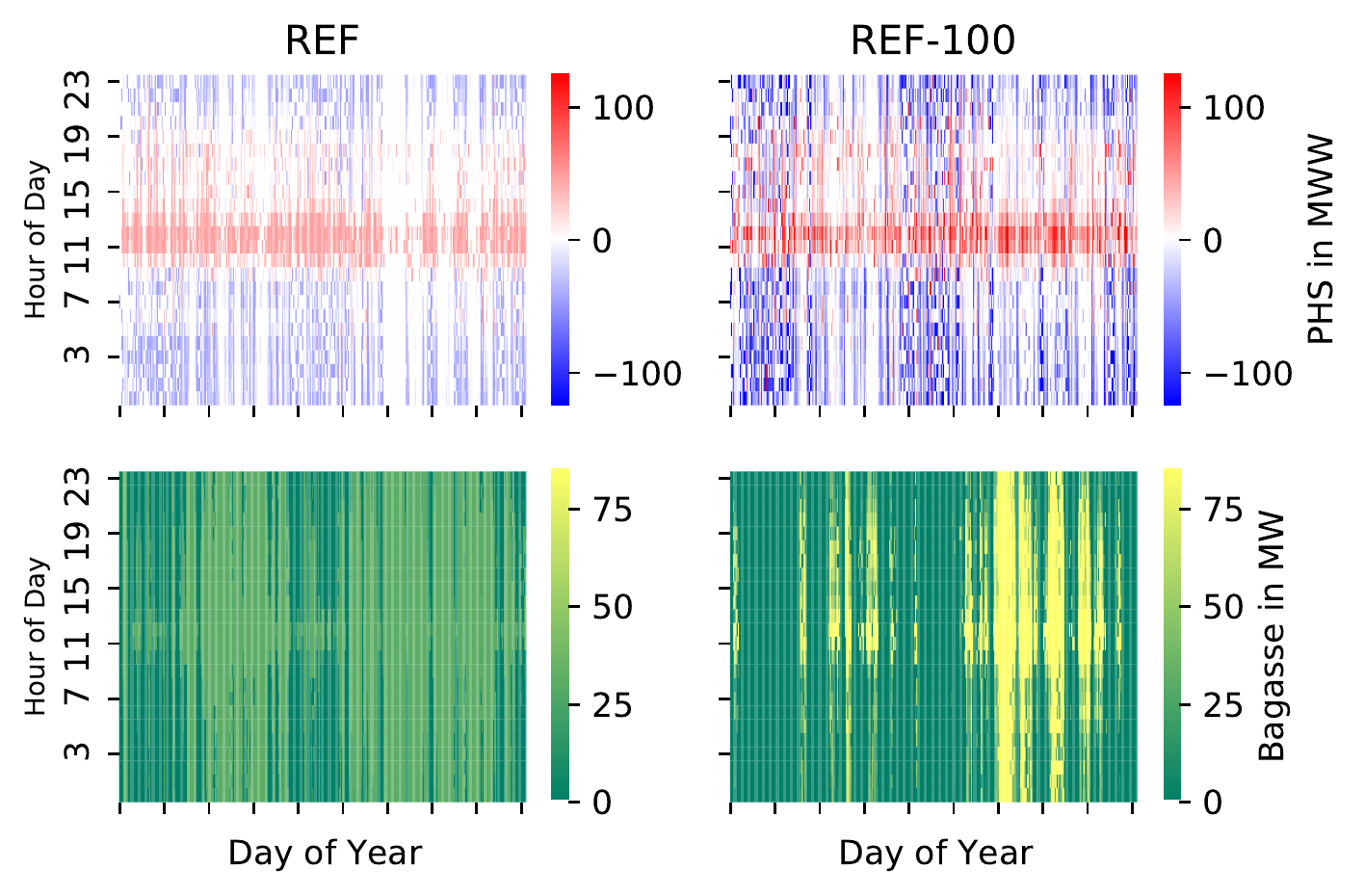}
    \caption{Heatmap for the operation of PHS (top) and bagasse (bottom) in the REF (left) and the REF-100 (right) scenario. }
    \label{fig:operation}
\end{figure}

To analyse the system operation, PHS storage and bagasse dispatch is shown in Figure \ref{fig:operation} for the REF and REF-100 scenarios. The PHS is mainly charged in the afternoon, due to the high PV supply, shifting energy into the night and morning. A seasonal shift between the month with lower wind supply can be observed. While absolute PHS increases from about 46~MW in REF to over 124~MW in REF-100, the pattern remains the same. In contrast, both the maximum supply and the pattern of bagasse change from the cost-optimal REF to the 100\% REF-100 scenario. In the REF scenario, bagasse operates in full load during many hours of the year, while in the 100~\% scenario, bagasse operates as a peaking and back-up component in the system to supply electricity during periods of low RE supply (e.g. when wind drops and PV is not available). Bagasse is less utilised in the summer months with high PV and wind supply and lower demand due to the absence of cruise ships, but also at the beginning of the year with high wind supply.

The operation of the dispatchable components underlines the need for a highly flexible operation to incorporate and manage long periods of no production, as evident in the drop of wind electricity generation, and full-load production in other periods. 

\subsection{Costs}
\begin{figure}[!h]
    \centering
    \includegraphics[width=\textwidth]{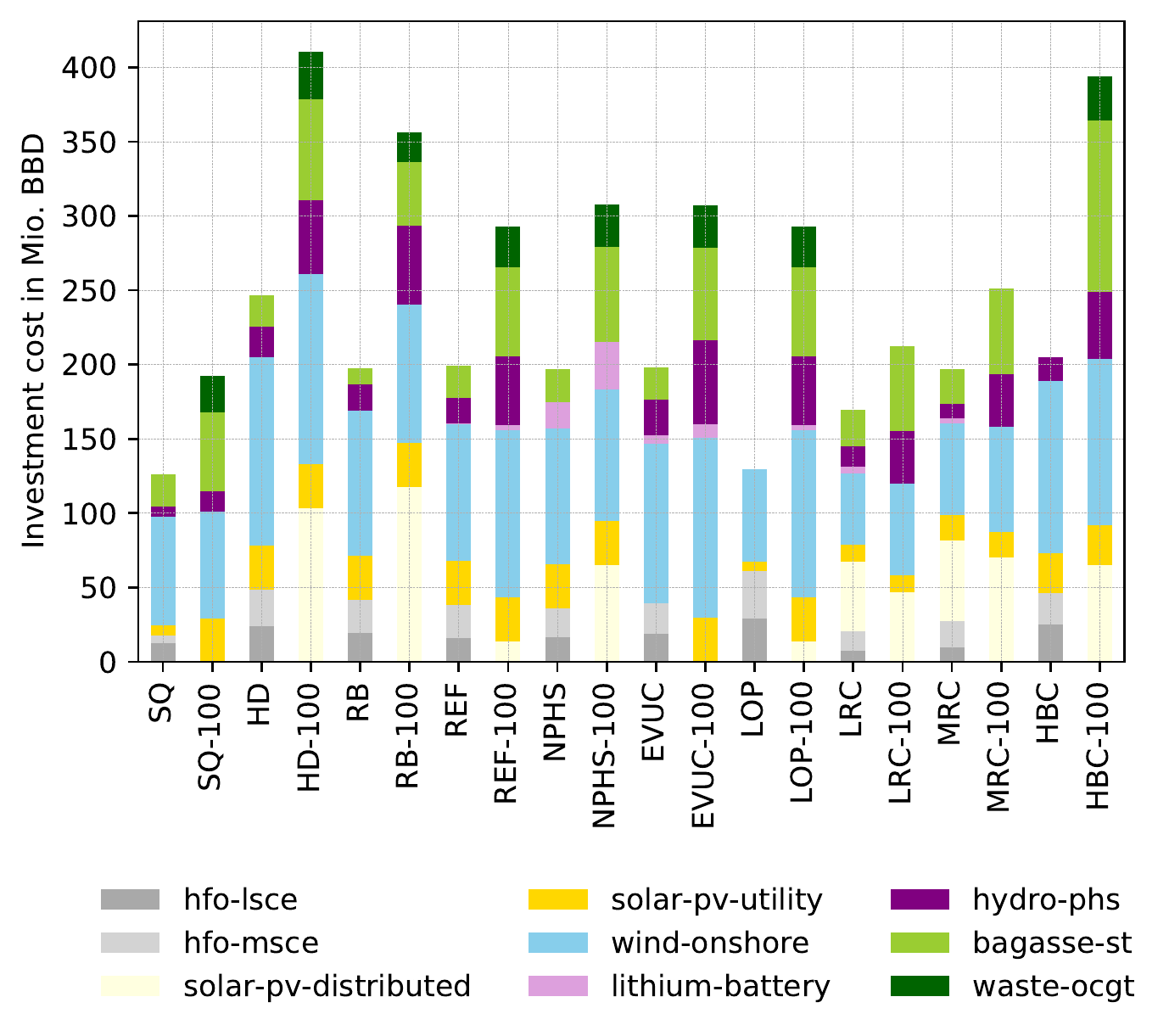}
    \caption{Investment cost per technology for all scenarios in the cost-optimal and the 100~\% renewable energy setup.}
    \label{fig:investment_cost}
\end{figure}

Investment costs play a crucial role, due to the Barbadian economic challenges, such as debt accumulation due to increasing and volatile prices of fossil fuels, limited sources of foreign exchange and a heavy dependence on external aid and donors. Figure \ref{fig:investment_cost} shows the annualised investment cost per technology for all scenarios. 100~\% RE systems typically require higher investment cost than the cost-optimal systems. Total annualised investment costs range from 126 Mio.\,BBD in the SQ to 204 Mio.\,BBD in the HBC (247 Mio.\,BBD in HD) in the cost-optimal cases and from 192 Mio.\,BBD (SQ) to 393 Mio.\,BBD (HBC-100) (410 Mio.\,BBD in HD-100) in the 100~\% RE scenarios. The costs range from 199 Mio.\,BBD to 293 Mio.\,BBD in the REF respective REF-100. It must be noted, that operational costs are significantly lower in the 100~\% RE systems due to the low or zero marginal cost of wind and PV, as well as lower cost for bagasse compared to heavy fuel oil, expect within HBC-100. While the share of energy provided from dispatchable renewable energies and storage is low, compared to the energy provided by volatile resources such as wind and photovoltaics, their share in the total investment costs is significantly high in the scenarios with 100~\% renewable energies. Lower RE costs, as in the LRC-100 and MRC-100 scenarios, lead to no investment in waste, whereas all other 100~\% scenarios rely on a waste investment. The increase in bagasse investment costs (HBC) leads to no investment in bagasse in the cost-optimal case, while in the HBC-100 scenario, higher investments are made in PV plants compared to the REF-100 scenario to compensate for the higher investment costs.

One main driver for installed capacities and associated required capital is the demand, which is demonstrated by comparing the investment costs of the REF and the HD scenario. Another main driver are the costs required for flexibility, which is indicated by high investment costs for the HBC-100 scenario compared to the REF-100 scenario.

\subsubsection*{LCOE and impact of financing costs (wacc)}
\begin{figure}[!h]
    \centering
    \includegraphics[width=\textwidth]{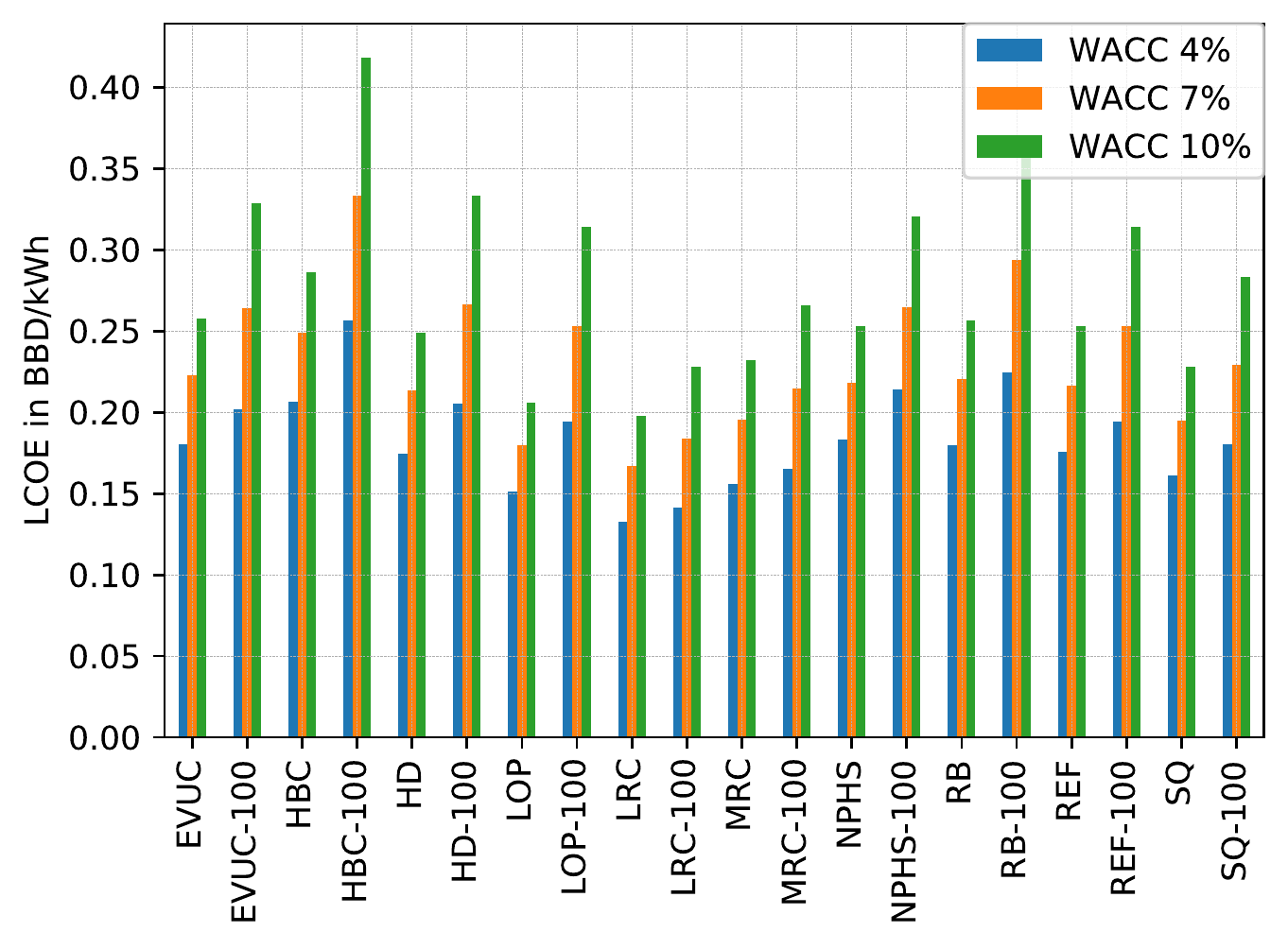}
    \caption{System LCOE for all scenarios in the cost-optimal case (COPT) and the 100~\% renewable energy case (100RE) in BBD per kWh with wacc of 4\%, 7\% and 10\%.}
    \label{fig:lcoe}
\end{figure}

LCOE have been calculated with three different assumptions for wacc (low 4\%, base 7\% and high 10\%). For all cases, the resulting systems' levelised cost of electricity (LCOE) show a moderate increase from cost-optimal to 100~\% RE systems (\autoref{fig:lcoe}). However, this increase is higher if renewable flexibility is expensive (HBC), restricted (RB) or the cost of fossil generation are low compared to renewable generation (LOP). 

With the base value of 7\% wacc, the lowest LCOE for the cost-optimal case occurs in the LOP scenario  (0.18~BBD/kWh), while the highest cost-optimal system LCOE can be identified for the HBC scenario with 0.25~BBD/kWh. In the 100\% RE scenarios, low LCOE of 0.18~BBD/kWh can be identified within the LRC scenario. In the reference scenario REF-100 costs are higher with 0.25~BBD/kWh but still significantly lower than for HBC-100 with 0.33~BBD/kWh. The general pattern is the same for all wacc assumption. However it is clearly visible, that with wacc of 4\% the LCOE decreases. On average the LCOE decrease by 22.6\% for 100\% RE cases and 18.2\% in COPT scenarios. In contrast, higher wacc of 10\% cause an average increase in LCOE of 24.21\% (100\% RE) and 16.5\% (COPT) respectively.\\

\begin{figure}[!h]
    \centering
    \includegraphics[width=\textwidth]{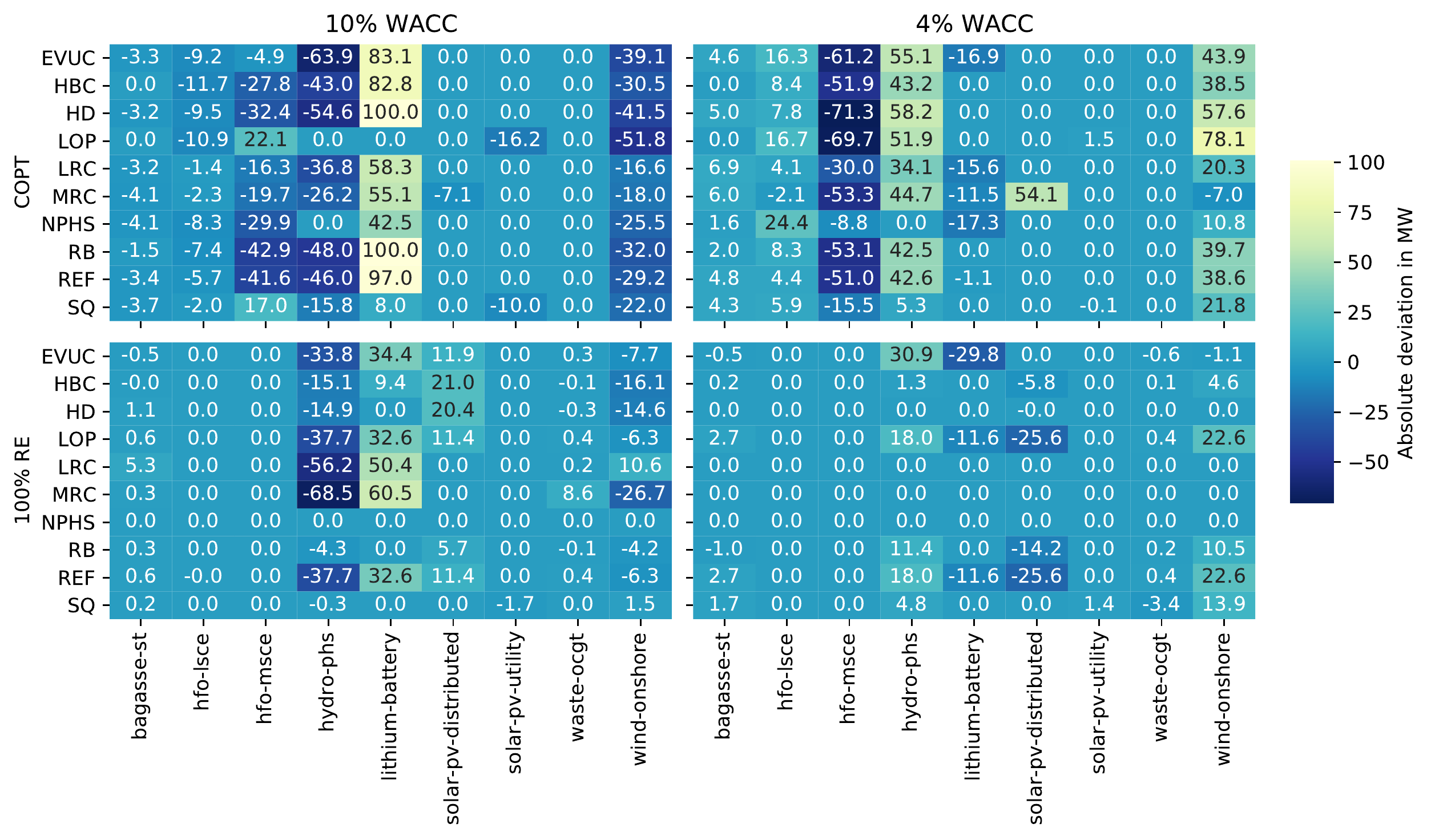}
    \caption{Deviation of installed capacities in MW with wacc of 10\% (right) and 4\% (left) compared to the reference case of 7\% for the cost optimal (COPT) and the 100\% RE case in all scenarios.}
    \label{fig:heatmap_wacc}
\end{figure}

Figure \ref{fig:heatmap_wacc} shows the impact of different wacc on the invested capacities. Generally, a varying wacc has a larger impact in COPT compared to 100\% RE scenarios. In the COPT case, lower wacc of 4\% causes a higher share of renewable energies, particularly noticeable in higher wind and PHS investment and decreased battery investment compared to the base case of a 7\% wacc.

Assuming a higher wacc of 10\%, the opposite effect becomes apparent. However, the NPHS scenario shows, that a higher wacc reduces investment in dispatchable units,  bagasse and hfo, and wind capacities, which is compensated by increased battery capacity. With low wacc of 4\%, fossil investment is reduced significantly in favour of higher PHS capacities. In a 100\% RE system, the high wacc causes a shift from wind to PV investment, with the opposite effect for a low wacc. Analysing storage options, with a high wacc investment in battery is favoured over PHS. The opposite effect, albeit not as drastic, is visible in some scenario settings with a low wacc. As the chosen 7\% wacc in the base case is already rather high, this underlines, that PHS is a robust solution on a way to 100\% RE for Barbados. The sensitivity analysis furthermore shows the importance of low and solid renewable energy financing options, especially for countries like Barbados.

\section{Discussion}
The results clearly indicate high potential shares of renewable energies in cost-optimal energy systems for 2030. The electrification of the cruise ship and transport sector is possible but requires additional investment capital, particularly for PHS and additional wind capacities. The high demand from electric vehicles reflected in the modelling results as well as the charging patterns pose a challenge for a future sustainable Barbadian energy system. If electric vehicle charging is uncontrolled, more storage is needed. This only leads to a slight increase in investment costs and LCOE. This underlines the role of demand response measures and energy policies to harness peak renewable energy generation. In the transport sector, a reduction in demand could be achieved by increasing the share of public transport and improving the infrastructure for (electric) bicycles to change the modal split. Additionally, an optimisation of the charging habits and the different effects controlled and uncontrolled charging have, needs to be further analysed and optimised, to counteract times with low renewable energy potential. This is highlighted within the modelling results, as higher RE shares are achieved within the REF scenario (controlled charging) compared to the EVUC scenario (uncontrolled charging). However, marginally higher investment costs are necessary within the REF scenario, respective the EVUC system. The present study suggests priority on the expansion of the wind capacity as it has a higher system value than PV. Whereas the technology LCOE for wind and PV are similar, a system with high wind capacities can achieve lower overall system LCOE. Wind power therefore forms an important pillar for the future renewable energy system for Barbados, regardless the seasonal variations within the wind resource, which are challenging.\\

Demand-side management (DSM) was not considered in this study. However, this can be seen as a valuable option for renewable energy integration, which should be explored in future analyses. As the model is based on oemof-solph \citep{krien_oemofsolphmodel_2020}, a DSM component can be included. First estimates of DSM Savings can be used from the Integrated Resource \& Resilience Plan for Barbados Activity B, Draft Report \citep{mott_macdonald_integrated_2021}. Future analysis can also include detailed grid planning, electricity planning and stronger economic analysis. These can underline and strengthen the arguments for 100~\% renewable energy.

\subsection*{Costs}
Estimates showed, that in 2013, fuel costs alone made up 73~\% (0.413 BBD/kWh) of the total electricity production costs (0.566 BBD/kWh) in Barbados \citep{hohmeyer_economic_2017}. In 2015, the purchase of international oil cost the Barbadian economy 377 million BBD, resulting in high electricity costs passed to the public in the form of the Fuel Clause Adjustment \citep{hohmeyer_100_2015}. A higher renewable energy share does not necessarily result in higher electricity costs. With the likely future reduction in RE cost, as modelled in the LRC and MRC scenarios, a RE share of more than 90~\% is possible, while keeping system LCOE below 0.20 BBD/kWh (for LRC even in the 100~\% scenario). Reaching a share of 100~\% RE comes with only a slight increase in LCOE from 0.18 to 0.20 BBD/kWh (REF). However, these values do not reflect integration costs, such as grid and balancing costs or taxes, but solely system generation costs. This needs to be considered and analysed in further detail. Regardless, the respective required (annual) investment costs of the HD scenario are 47.6 Mio.\,BBD per year higher than in the REF scenario. These cost can be interpreted to be available for a 40~\% demand reduction. Although the LCOE may not differ largely between the HD and REF scenario, higher capital investment is required. As supported in previous studies, there is a sharp increase in investment costs as the energy system approaches a 100~\% RE. By assuming a higher than usual lifetime and nominal efficiencies of new technologies, is it feasible to built newest available technologies or built older, cheaper plants, they will be running in part-load anyway so they won't cover the nominal efficiencies, however this means that conservative cost optimal with regard to 100\% systems, intertwined with grid etc. 

\subsection*{Flexibility}
Lack of or expensive flexibility within an energy system can lead to significant problems, especially when demand increases, the flexibility potential is restricted. In Barbados, flexibility is provided by biomass (bagasse) and waste and will play a significant role for investment costs and LCOE in the future energy system. Further research is necessary to determine how much of the currently installed capacity might be used as flexible generation units. In addition, the use of biomass may be limited due to operational requirements in systems with high shares of RE.

\subsection*{Limitation of Study}
The results of this study have to be critically read in the light of the used method and model. Within the model supply units are aggregated and not represented as individual units. In addition, grid constraints and spinning reserves are not considered. Another important restriction of the model is the applied perfect-foresight approach, which particularly matters within the context of the specific Barbadian weather pattern. The drop in wind supply in autumn in combination with restricted biomass resources and storage capacities requires long-term planning of dispatchable power plant operation and storage management. While the perfect-foresight approach allows to optimise this operation for a whole year, it will be much more challenging in reality. The maximum installed capacities are constrained, due to meteorological and geographical conditions. The wind resource was capped at 472~MW in keeping with the recommendations of \cite{rogers_desktop_2017}. The solar resource is constrained at 234~MW at the residential level. Without studies to support the expansion of utility scale solar on the island, this potential was constrained to 80~MW for all scenarios. Considering utility-scale PV is the cheapest renewable resource and was always selected by the model and used at maximum potential, it is vital to verify the assumed potential.

The assumptions made, may lead to rather optimistic results in terms of capacities and costs. The costs for the integration of renewable energies, such as grid expansion and back-up capacities for stable system operation, must be added to the costs presented in this study. While integration costs are highly context specific literature values suggest that these costs can range of 35 to 50~\% of the total generation costs \cite{hirth_integration_2015}. The challenges of long-term operational planning under resource constraints in 100~\% RE systems also need to be considered.

\subsection*{Value of open source}
The boundaries and specifications posed within the context of Barbados, such as load profiles and possible resource capacities are key factors when analysing electricity systems. This information, however, is not always available to the public nor to scientists. Additionally, future economic and political developments can change fundamentally within a short period of time altering assumptions taken within the model. The applied Open Energy Modelling Framework offers the possibility to flexibly adjust input data and thus quickly and transparently adapt to changing circumstances without reducing the traceability of results.

\subsection*{Shore-to-ship power}
The study describes shore-to-ship demand based on the annual arrival of cruises ships and varies with the tourist season over the course of the year. The cruise ship demand contributes significantly to the total energy demand of Barbados. The cruise ship demand pattern is almost complementary to the wind generation profile, which further supports the suitability of this resource for the Barbadian energy system. However, more research examining how renewable energy sources can optimally meet the daily cruise ship demand through e.g. scheduled docking times by the system operator, would be beneficial. 

\section{Conclusion}
Due to good renewable energy resources and future cost developments, a cost-optimal system design in Barbados already features a share of over 80~\% renewable energy, assuming a future reduction in RE costs with current oil prices. Even with increased demand due the electrification of cruise ships and passenger transport, a renewable energy share of over 80~\% is achieved, with only slightly increased system LCOE (0.22 BBD/kWh respective 0.20 BBD/kWh). Particularly the energy potentials of waste and bagasse and the flexible operation of these systems need to be validated to ensure a stable 100~\% RE-system for the future. In the case of Barbados, imported bio-diesel for backup may be required due to the specific wind and solar pattern. Nonetheless, PHS can lower the costs of a 100~\% renewable set-up, additionally reducing the consumption of resources needed for batteries. The value of pumped hydro is only slightly affected by higher costs of renewable energy sources, lower oil prices or a demand reduction. Therefore, when aiming for 100~\% RE, the results indicate that PHS is a no-regret option for the Barbadian energy systems design.
An open source energy system model as introduced here for the island of Barbados marks an important step towards decarbonising SIDS. By using an open source model, the problems of external donor agencies and the possibility of biased results, especially in the context of Small Island Developing States, are addressed. The study additionally underlines the feasibility of RE systems, even with an increasing demand including the electrification of sectors other than electricity generation.
\clearpage

\small\textbf{Data availability}\\
\footnotesize
The model is available at \url{https://github.com/znes/oemof-barbados/releases/tag/v0.2}.
\\

\small\textbf{CRediT authorship contribution statement}\\
\footnotesize
A. Harewood: Conceptualization, Resources, Data curation, Validation, Formal analysis, Investigation, Supervision, Project administration, Writing - original draft. F. Dettner: Conceptualization, Resources, Data curation, Validation, Formal analysis, Investigation, Methodology, Writing – original draft, Writing – review \& editing. S. Hilpert: Conceptualization, Software, Methodology, Investigation, Visualization, Writing – original draft.
\\

\small\textbf{Declaration of competing interest}\\
\footnotesize
The authors declare no known conflict of interest, financial or otherwise.
\\

\small\textbf{Acknowledgements}\\
\footnotesize
André Harewood and therefore his work is supported by the National Development Scholarship Scheme Barbados by the Ministry of Education, Technological and Vocational Training.

\clearpage

\appendix

\begin{landscape}
\section{Scenario studies summary}
\begin{table}[!h]
\centering
\footnotesize
\caption{Overview of key values from scenario studies on renewable energy in Barbados, current as of study year}
\label{tab:overview_of_pre_studies}
\begin{tabular}{p{3cm}p{3.5cm}p{3.5cm}p{3.5cm}p{3.5cm}}
\toprule 
 & \cite{blp_integrated_2014}  & \cite{irena_barbados_2016} & \cite{hohmeyer_economic_2017} & \cite{gob_barbados_2019}\\ \midrule 
Curr. capacity      & 239.1 MW\textsubscript{2012}              & 241.5 MW\textsubscript{2015}  & 240 MW & n/a \\
Exp. capacity       & 293.3 MW\textsubscript{2036}              & 450 MW\textsubscript{2030}    & 395 MW & n/a \\
Curr. demand        & 980 GWh/a\textsubscript{2012}             & n/a & 912 GWh/a\textsubscript{2013} & 11,297 BOE per day \\
Exp. demand         & 903\,-\,1,986 GWh/a\textsubscript{2036}   & 998 GWh/a\textsubscript{2030}& 1350 GWh/a & n/a \\
Curr. generation    & n/a                                       & 1092 GWh/a\textsubscript{2015} & 970 GWh/a\textsubscript{2013} & n/a \\
Exp. generation     & n/a                                       & 998 GWh/a\textsubscript{2030}& n/a & 1600 GWh \\ 
Curr. peak demand   & 163 MW\textsubscript{2011}                & 158 MW\textsubscript{2015} & 150 MW & n/a \\
Exp. peak demand    & 208.1 MW\textsubscript{2011}              & 145 MW\textsubscript{2030} & 140-300 MW\textsubscript{2036} & n/a\\
Exp. share of RE    & 1.2 - 29\%                                & max. 76\% & 100\% & 76\% \\
Storage             & wind with 10\% battery           & 150 MW battery\textsubscript{2035}             & 3 GWh\textsubscript{2035}  PHS & - \\
Biomass             & 25 MW                                     & 18 MW (for 100\% 54 MW)   & 25-40 MW (35 GWh) & 39 MW\textsubscript{2035} \\
Waste potential     & 60 MW \textsubscript{2035}                                     & 2.2 MW & 11 MW & 40 MW (WtE)\\
Solar (PV)          & -                                         & 60 MW\textsubscript{2030} & 219 MW - 265 MW\textsubscript{2035}& 195 MW\textsubscript{2037}\\
Wind                & -                                         & 15 MW\textsubscript{2030} & 219 - 265 MW\textsubscript{2035} & 127 MW\textsubscript{2037}\\
Natural Gas         & n/a                                       & n/a & n/a & 49 MW\textsubscript{2037}\\
Electrification rate vehicles  & -                                         & 20\% -50\% EV & 100\% EV & 100\% EV \\
Cruise ship demand       & n/a                                       & n/a & n/a & n/a \\
\bottomrule
\end{tabular}
\end{table}
\end{landscape}

\section{Mathematical symbols}

\begin{table}[!h]
    \centering
    \caption{Sets used in the model description and values of sets used within the applied scenarios.}
    \begin{tabular}{llp{3cm}p{4cm}l}
       \toprule
        Symbol & Index & Description & Elements of sets in scenarios & Unit\\
        \midrule
        \(T\) & \(t\) & Time steps & \{1...8760\} & h\\
        \(V\) & \(v\) & Volatile renewable units & \{wind-onshore, pv-utility, pv-distributed\}& MW\\
         \(D\) & \(d\) & Dispatchable renewable units & \{bagasse-st, waste-ocgt\}& MW\\
        \(C\) & \(c\) & Conventional units& \{hfo-lsce, hfo-msce\}& MW\\
         \(L\) & \(l\) & Load types & \{cruise-ship, ev, el-demand\}& MW\\
        \(S\) & \(s\) & Storage units & \{lithium-battery, PHS\} & MW, MWh\\
        \(U\) & \(u\) & All supply units & $R \cup C \cup S \cup D$  & -\\
        \bottomrule
    \end{tabular}
    \label{tab:sets}
\end{table}

\begin{table}[!h]
    \centering
    \caption{Optimisation variables used in the model description.}
    \begin{tabular}{lll}
       \toprule
        Symbol & Description\\
        \midrule
        \(\textbf{e}_{s,t}\) & Storage level (energy) of storage \(s\) at time step t\\
        \(\textbf{e}^{nom}_s\) & Nominal storage level (energy of full storage) of storage $s$\\
        \(\textbf{h}_{t}\) & Fuel consumption at time step \(t\)\\
        \(\textbf{p}_{t}\) & Power output/input  at time step \(t\)\\
        \(\textbf{p}^{nom}\) & Upper limit of power output\\
        \(\textbf{p}^{excess}_t\) & Excess variable\\
        \bottomrule
    \end{tabular}
    \label{tab:variables}
\end{table}

\begin{table}[!h]
    \centering
    \caption{Exogenous model variables used in the model description.}
    \begin{tabular}{lll}
      \toprule
Symbol & Description\\
\midrule
\(\tau\) & Lenght of time step (in hours)\\
\(\overline{p}_{i}\) & Upper power investment limit of unit \(i\)\\
\(\underline{p}_{i}\) & Lower power investment limit of unit \(i\)\\
\(\overline{e}_{s}\) & Upper energy investment limit of storage $s$\\
\(d_{l, t}\) & Electricity demand of type $l$ at time step $t$\\
\(d^{peak}\) & Total electricity peak demand within the year\\
\(\overline{h}_{d}\) & Maximum fuel consumption of unit d\\
\(\eta^{loss}_s\) & Standing loss of storage $s$\\
\(\eta^{in}_s\) & Charge efficiency of storage $s$\\
\(\eta^{out}_s\) & Discharge efficiency of storage $s$\\
\(c^{opex}_u\) & Operational expenditure of unit $u$\\
\(c^{capex,p}_i\) & (Annualised) power expenditure of unit $i$\\
\(c^{capex,e}_s\) & (Annualised) energy capital expenditure of storage $s$\\
\(c^{profile}_v\) & Generation profile of volatile renewable energy unit $v$\\
\(RE^{share}_v\) & Renewable energy share between 0 and 1.\\
\bottomrule
    \end{tabular}
    \label{tab:parameters}
\end{table}

\clearpage

\section{Cruise demand}
\begin{figure}[!h]
    \centering
    \includegraphics[width=0.85\textwidth]{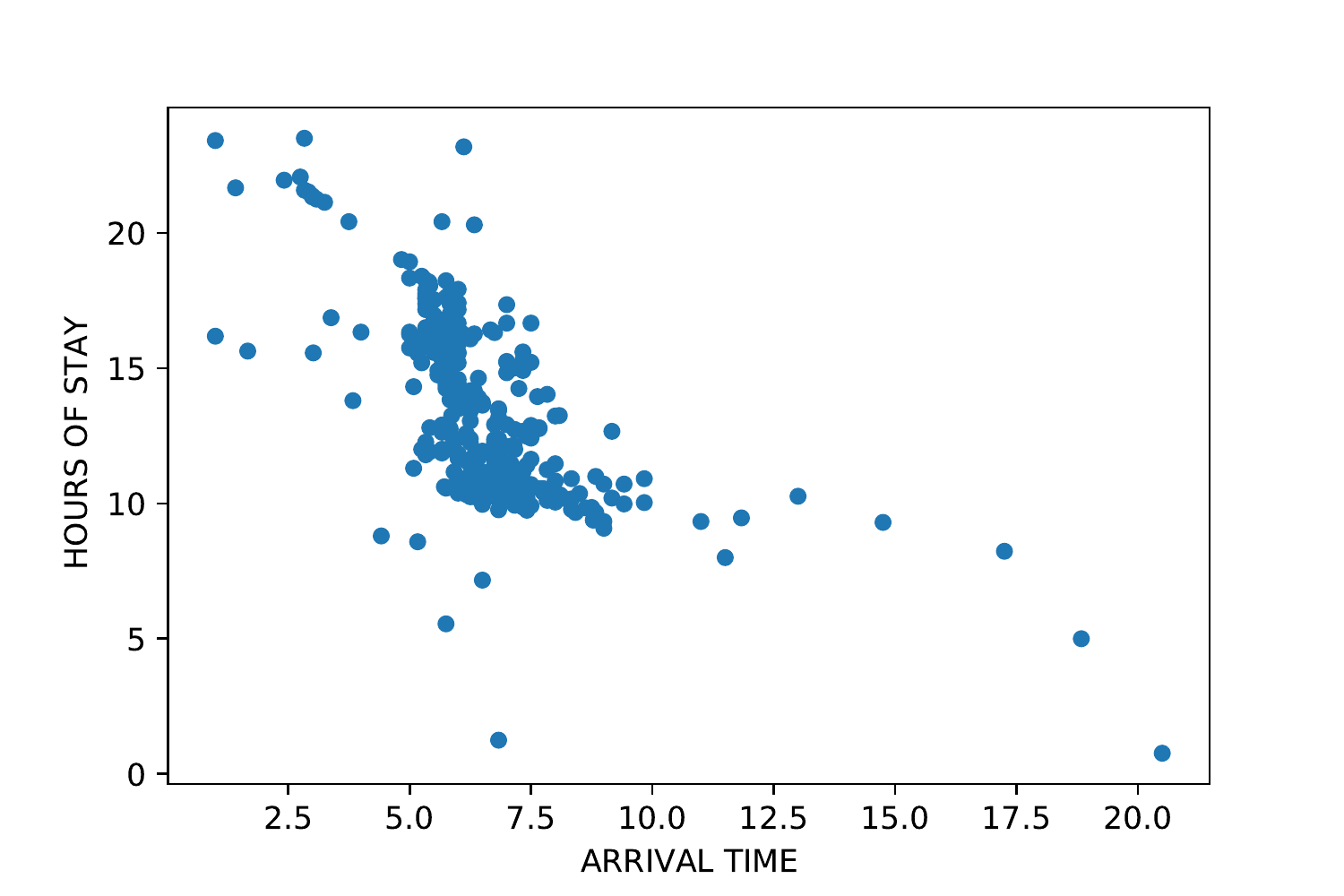}
    \caption{Arrival times and duration of stay of cruise ships in the port of Barbados for 2018 (analysis based on \cite{bpi_port_2020})}.
    \label{fig:arrival_and_stay}
\end{figure}

\begin{figure}[!h]
    \centering
    \includegraphics[width=0.85\textwidth]{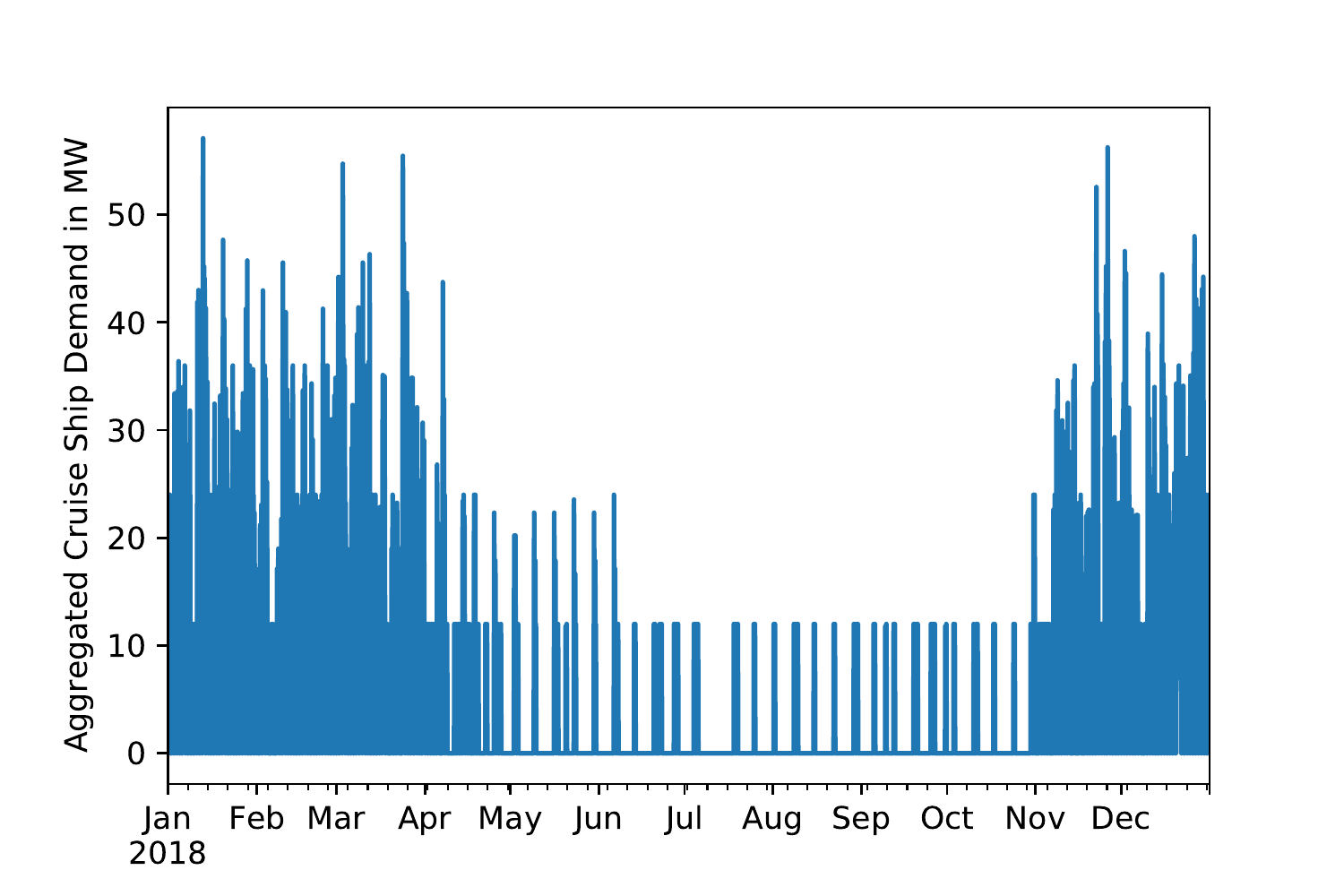}
    \caption{Cruise demand for 2030, modelled from detailed port statistics based on actual arrivals and duration of stay of 2018 (analysis based on \cite{bpi_port_2020}).}
    \label{fig:cruise_demand}
\end{figure}

\clearpage

\section*{Supplementary Material: Open source modelling of scenarios for a 100\% renewable energy system in Barbados incorporating shore-to-ship power and electric vehicles}

\subsection*{Background Barbados}

Barbados is defined as a small island developing state (SIDS) and therefore has some unique characteristics. The UN Conference on Environment and Development in 1992 provided the first definition for this distinct group of developing nations geographically located in the Caribbean, Atlantic, Indian and Pacific Oceans \citep{robinson_adapting_2018}. They are characterised by small geographical areas, insularity, remoteness and are prone to natural disasters as well as the negative impacts of climate change \citep{pelling_small_2001}. The small island character creates limited markets, often with erratic domestic revenues, which are further exacerbated by high costs of public services \citep{briguglio_small_1995}. Many SIDS experience heavy fiscal burdens, limited sources of foreign exchange, dept accumulation and a heavy dependency on external aid and donors, while being almost 100~\% reliant on fossil fuel imports \citep{atteridge_development_2019}. Another crucial challenge is the financial dependence on bilateral or multinational loans \citep{oecd_making_2018}. These factors, along with the ubiquitous pressures of global warming and the need for CO\textsubscript{2} mitigation, urge SIDS to redirect fossil fuel spending to other critical areas. This twin challenge could be addressed through a sustainable transformation of the energy sector. According to \cite{robinson_adapting_2018}, the main incentive behind a shift from (imported) fossil to (local) renewable energy sources is economic survival, rather than climate change mitigation.

\subsection*{Energy system}
The Barbadian electricity system relies almost entirely on imported fossil fuels. In 2020, 95.4 \% came from oil and diesel, with heavy fuel oil as the primary fuel source, followed by kerosene and diesel \citep{healey_energy_2020}. There is a total of three conventional generation sites referred to as Spring Garden (153.1 MW), Garrison Hill (13 MW) and Seawall (73 MW) \citep{blp_integrated_2014}. Fuel as input for generation changes depending on the annual oil price and the subsidy structure \citep{gob_barbados_2019}. Spring Garden therefore uses heavy fuel oil, whereas the gas turbine at Garrison Hill and Seawell are operated on diesel fuel \citep{espinasa_achieving_2016}. \autoref{tab:powerplants} summarises the installed power plants as well as the proposed decommissioning dates. The total installed capacity in 2018 was therefore 286.6 MW, with a share of 10.5 \% of installed renewable energy in the form of photovoltaics (PV) \citep{irena_renewable_2019}. In 2015 only 2 \% were supplied by renewable energy \citep{gob_barbados_2019}. There is a 100 \% access rate to electricity and Barbados continues to have a reliable power supply with fewer blackouts than other regional territories according to \cite{espinasa_achieving_2016}. The energy demand by sectors breaks down to 41 \% electricity generation, 27 \% transportation, 10 \% commercial and public, 9 \% industry, 9 \% residential and 5 \% other \citep{irena_barbados_2016}. The retail energy price in Barbados, which can rank among the highest in the Caribbean, varies significantly with the price of international oil and is allocated to energy consumers through the Fuel Clause Adjustment \citep{gob_barbados_2019, henry_key_2015}.

\begin{table}[!ht]
    \centering
    \footnotesize
    \begin{tabular}{lllll}
    \toprule
    Power station  & Fuel & Capacity & Generation type & Proposed decommission \\
    \midrule
    Spring garden      & HFO  & 40 MW    & Steam turbine   & 2017\\
    Spring garden      & HFO  & 50 MW    & LS Diesel generation       & 2019\\
    Spring garden      & HFO  & 3.7 MW   & Co-generation unit   & 2019\\
    Spring garden      & HFO  & 59.4 MW  & LS Diesel generation   & 2036\\
    Seawell            &Diesel & 73 MW   & Gas turbine  & 2022-2028\\
    Garrison Hill      & Diesel & 13 MW  & Gas turbine  & 2022\\
    \bottomrule
    \end{tabular}
    \caption{Fossil fuel power plants in Barbados, according to \citep{espinasa_achieving_2016}, \citep{gob_barbados_2019} and \citep{blp_integrated_2014} with proposed decommissioning dates. As of December 2021, all plants continue to be active.}
    \label{tab:powerplants}
\end{table}

The peak demand and total generation were examined in several studies solely as formal statistics as the figures are not released to the public. Between the period of 1993 to 2013, the peak demand grew from 92 to 152 MW and reached a maximum of 167.5 MW in 2012 \citep{espinasa_achieving_2016}. Between 2014 and 2019, the peak demand oscillated between 150 and 155 MW, which fits the \citep{irena_barbados_2016} estimate of 158 MW peak demand in 2015. In 2018 the peak demand were 152.3 MW, which corresponded to a total generation of 996 GWh/a \citep{healey_energy_2020}. The slight drop in the peak demand within recent years can be attributed to limited economic growth and the expansion of distributed PV generation \citep{espinasa_achieving_2016}.  \cite{hohmeyer_economic_2017} puts the total demand in 2013 at 912 GWh/a with a total production of 970 GWh/a. Accordingly, \cite{irena_barbados_2016} reports a generation of 1090 GWh in 2015. Official government estimates show that in 2019 and 2020 the total generation was 943 GWh and 893 GWh per year respectively. The significant drop can be attributed to a rapid contraction in economic activity, specifically, within the tourism industry due to the COVID-19 pandemic \citep{research_department_energy_2020}.

\subsubsection*{Wind and Solar}
Weather conditions in Barbados are promising for a cost efficient RES. According to \cite{alleyne_100_2014}, the wind resource is steady, constant and reliable in direction due to the trade winds. The wind speeds range between 4.8 and 8.0 m/s at 10 m hub height. For potential wind energy developments, based on data from 2000 to 2019, an average capacity factor of 36.97 \% was calculated. However, during this period, the average capacity factor of all lowest months in each year is only 19.7 \%. The reason for this difference is the specific characteristic within the temporal dynamic of wind, which drops in the months September and October. The dynamic is shown in Figure \ref{fig:capacity-factor} based on selected weather years. \cite{hohmeyer_100_2015} states a theoretical potential of 4.3 GW which needs to be reduced to a technical potential of around 600 MW for wind energy deployments. The Government of Barbados \citep{gob_barbados_2019} state a maximum of 150 MW wind onshore and 150 MW wind offshore for 2030. \cite{rogers_desktop_2017} conducted a study using WindFarmer Pro analysing the onshore wind resources in Barbados, recommending a total of 472 MW of installed wind capacity. The (average) daily solar capacity factor is significantly lower but more constant over the course of the year compared to the capacity factor for wind. \cite{hohmeyer_economic_2017} estimates an average solar irradiance of 2,196 kWh/m\textsuperscript{2}, which peaks from April to August and falls of slightly from August to December \citep{alleyne_100_2014}. This trend is in keeping with the rainy season that would be characterised by partly cloudy skies and a slight reduction in the solar irradiance \citep{field_effect_2015}. Presently, over 95 \% of the installed renewable energy grid-connected systems on Barbados are PV installations \citep{caricom_energy_2018}, which are about 40 MW of distributed PV, including a 10 MW solar farm in the north of the island. \cite{irena_barbados_2016} estimates the solar energy generation in 2017 at 30 GWh. The total solar potential for 2030, estimated by \cite{gob_barbados_2019} is split into 205 MW solar centralised and 105 MW solar decentralised. \cite{hohmeyer_100_2015} states a theoretical potential of 200 MW, requiring an area of 1.6 km\textsuperscript{2}.

\begin{figure}[!h]
    \centering
    \includegraphics[width=0.9\textwidth]{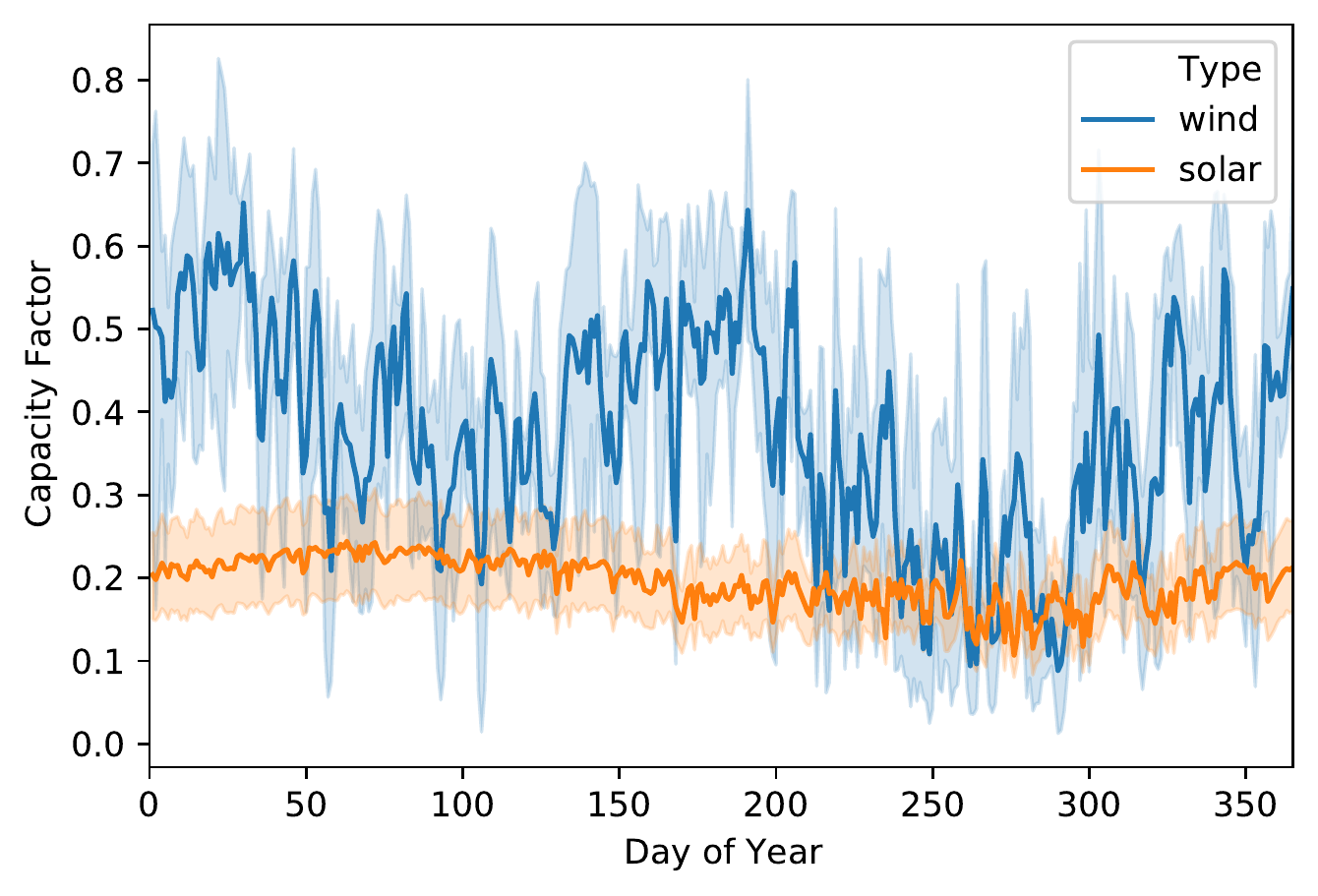}
    \caption{Average wind and solar daily capacity factor and 95 \% confidence interval of four years (2002, 2004, 2010, 2014) based on \emph{Renewables.Ninja} \citep{pfenninger_long-term_2016} data for location at 13.32 latitude and -59.6321 longitude.}
    \label{fig:capacity-factor}
\end{figure}

\subsubsection*{Bioenergy}
The relevance of bioenergy for a 100 \% RES in Barbados to complement the wind and solar potential was stated inter alia within \cite{espinasa_achieving_2016, gob_barbados_2019} and \cite{irena_barbados_2016}. Considering that the island has an extensive natural gas network, that supplies one million cubic feet per day to the public, the Government of Barbados has recognised the possibility of supplementing the natural gas mix with biogas \citep{gob_barbados_2019}. Historically, Barbados has always been a high-cost sugar producer, due to the lack of economies of scale and costs associated with growing sugar cane locally \citep{mitchell_sugar_2005}. Therefore, bagasse as a residue from the sugar production process can be used for the generation of bioenergy. \cite{marshall_bio-digestion_2019} confirmed that 1.5 million m\textsuperscript{3} of biogas could be produced from guinea grass, through anaerobic digestion. \cite{irena_renewable_2019} estimates a biomass potential of 54 MW. All analysed studies agree, that the determinant factor is the available area for biomass. According to \cite{undp_preliminary_2009} Barbados had 8,000 ha under sugar cane cultivation as of 2010. The total arable area was estimated at 19,000 ha, whereas 30 MW of production capacity could be met with bagasse, cultivated on 1,610 to just over 2,000 ha.
 
The island has a significant problem regarding the management of waste as approximately 900-1000 tonnes are sent to the landfill, of which about 300-500 tonnes are municipal solid waste (MSW), that could technically be used for waste-to-energy recovery \citep{marshall_bio-digestion_2019,rj_burnside_international_preliminary_2014,simmons_and_associates_barabados_2015}. However, these estimates have not been supported by large-scale feasibility studies or pilot projects. In 2014, a 3 MW waste-heat recovery plant was installed and \cite{irena_barbados_2016} assumes a possible production of 10.57 GWh in 2030. \cite{hohmeyer_economic_2017} estimates a higher possible capacity of 11 MW solid waste combustion in 2030. The \cite{undp_preliminary_2009} analysis estimates around 365,000 tonnes of raw material to be used in a production capacity of 3-5 MW.

\subsection*{Cruise Ships \& Electric Vehicles}
The electrification of sectors, which are traditionally powered by fossil fuels is being pursued. According to \cite{schuhmann_visitors_2019}, tourism had a total contribution of almost 40 \% to the GDP as well as the national employment in 2015. In 2018, more than 800,000 cruise ship passengers were recorded in Barbados \citep{bpi_port_2020}. The cruise ship industry is a substantial source of noise and air pollution, emitting NO\textsubscript{X}, SO\textsubscript{X}, GHG and particulate matter emissions, especially when the ships are berthed in the harbour \citep{lind_potential_2018, endresen_emission_2003}. The increased health burden from the resulting air pollution can lead to increased cases of illnesses such as asthma and respiratory infections, which burden the health system and the well-being of the population in the country with additional costs (for further reading, please refer to inter alia \citep{zis_evaluation_2014, tang_impact_2020, viana_estimated_2020, rabl_public_2000}). A more sustainable alternative are shore-to-ship power connections, which use the onshore power grid to meet the berthing power demand of ships. This concept has the potential to reduce CO\textsubscript{2} and air pollutant emissions substantially. Considering the sector's importance for economic development and as a regular seasonal energy demand, it is vital to be incorporated into energy related studies and energy system modelling. However, according to \cite{irena_barbados_2016}, cruise ship companies have expressed their doubt and reluctance to use such shore-to-ship power connections. Also, the Government of Barbados \citep{gob_barbados_2019} doubts the possibility to electrify cruise tourism, arguing, that resulting GHG emissions are not attributable to the host country alone. 

As an island system aiming for 100 \% renewable energy and due to air quality considerations, it is apparent to include the passenger transport sector in the modelling approaches. According to \cite{irena_barbados_2016} 81.5 \% of all registered vehicles are privately owned passenger vehicles, which could easily be replaced by similar classed electric vehicles. 

\subsection*{Climate Change and Energy Policy}
As most SIDS, Barbados is highly vulnerable to climate change as the atmospheric temperature increases, sea levels rise and weather patterns change \citep{undp_barbados_2020}. The island has therefore not only committed to several overall sustainable development goals as a member of the SID Group of Nations, but has also developed a Nationally Determined Contribution to reduce greenhouse gas emissions. Specifically, this includes the formulation of a Nationally Appropriate Mitigation Action (NAMA) as described in the Sustainable Energy Framework (SEF 2010), aimed at connecting the mitigation of climate change to a broader sustainable development for the energy sector. At the same time, Barbados's contribution to anthropogenic GHG emissions have been minuscule in relative and absolute terms. With regard to CO\textsubscript{2} emissions, controversial values are stated within the literature. A study by the government of Barbados \citep{gob_barbados_2019} puts per capita CO\textsubscript{2} emissions at 29.5 t in 2015, whereas in 2016, the total CO\textsubscript{2} emissions were about 1,276 kt, with 4.5 tCO\textsubscript{2}/capita according to \cite{macrotrends_barbados_2021}.

As the pursuit of higher shares of renewable energy sources is not primarily climate change related but also a facilitator to address economic and fiscal challenges caused by the high dependence on imported fossil fuels, the Government of Barbados has committed to transforming the entire energy sector, including electricity generation and transportation \citep{irena_barbados_2016, gob_barbados_2019}. Consequently, the most recently approved Barbados National Energy Policy (BNEP) for 2030 by the Ministry of Energy and Water Resources wants to diversify the energy sector, whilst aiming at a 100 \% renewable energy-based system in 2030 under the keywords diversity, efficiency, affordability and collaboration \citep{gob_barbados_2019}. The 2030 agenda calls for an elimination of the use of fossil fuels for the local transport sector, switching to bio fuels or increased electrification.

\subsection*{Users and Utility}
Within the context of 100 \% renewable energy scenarios for Barbados, the study shows that high electricity costs will warrant further consideration of suitable market tools for ensuring the lowest energy system costs. Selecting suitable market instruments such as feed-in-tariffs, renewable energy portfolio standards and auctions are becoming more critical moving forward. A more detailed market analysis that considers the unique challenges for SID states, such as insularity and small market size, will lend more insight regarding achieving greater sustainable development. The Government of Barbados has concluded that the Barbadian energy system lacked the level of competition and capable local market participants required for energy auctions \cite{gob_barbados_2019}. Simultaneously, the country intends to prioritise local investment in the energy sector without significantly favouring foreign investors. Similarly, renewable portfolio standards would require an energy market for green certificates to produce renewable energy, which is also limited by the small market size of Barbados that favours vertically integrated monopolies such as the local utility. Ultimately, a feed-in-tariff system is suggested as the best market instrument that fits the limitations of the Barbadian power system and provides stability as requested by domestic shareholders during the most recent dialogues within the energy sector \cite{mott_macdonald_integrated_2021}. Furthermore, the issue of financing the energy system becomes even more apparent as most of the capital for major energy infrastructure for the utility will be sourced from international donor agencies.

\clearpage

\bibliography{literature}

\begin{thebibliography}{53}
\expandafter\ifx\csname natexlab\endcsname\relax\def\natexlab#1{#1}\fi
\providecommand{\bibinfo}[2]{#2}
\ifx\xfnm\relax \def\xfnm[#1]{\unskip,\space#1}\fi
\bibitem[{Alleyne(2014)}]{alleyne_100_2014}
\bibinfo{author}{Alleyne, A.} (\bibinfo{year}{2014}).
\newblock {\it \bibinfo{title}{100\% {Renewable} {Electricity} {System} for
  {Barbados}: {The} {Solar} {Resource} in {Barbados}}\/}.
\newblock \bibinfo{type}{Final {Course} {Report}} \bibinfo{number}{RNEM 6010}
  The University of the West Indies Cave Hill Campus
  \bibinfo{address}{Bridgetown, Barbados}.
\bibitem[{Atteridge \& Savvidou(2019)}]{atteridge_development_2019}
\bibinfo{author}{Atteridge, A.}, \& \bibinfo{author}{Savvidou, G.}
  (\bibinfo{year}{2019}).
\newblock \bibinfo{title}{Development aid for energy in {Small} {Island}
  {Developing} {States}}.
\newblock {\it \bibinfo{journal}{Energy, Sustainability and Society}\/},  {\it
  \bibinfo{volume}{9}\/}, \bibinfo{pages}{10}.
\bibitem[{Barbadian-Government(2020)}]{research_department_energy_2020}
\bibinfo{author}{Barbadian-Government} (\bibinfo{year}{2020}).
\newblock {\it \bibinfo{title}{Energy {Bulletin} 2020 – {Energy}.gov.bb}\/}.
\newblock \bibinfo{type}{Energy {Bulletin}} \bibinfo{number}{January - December
  2020} Research Department \bibinfo{address}{Energy Divison}.
\bibitem[{Blechinger \& Richter(2014)}]{blechinger_barriers_2014}
\bibinfo{author}{Blechinger, P.}, \& \bibinfo{author}{Richter, K.}
  (\bibinfo{year}{2014}).
\newblock {\it \bibinfo{title}{Barriers and solutions to the development of
  renewable energy technologies for power generation on {Carribaen} island
  states}\/}.
\newblock \bibinfo{type}{Workshop {Presentation}} Reiner Lemoine Institut.
\bibitem[{{BL\&P}(2014)}]{blp_integrated_2014}
\bibinfo{author}{{BL\&P}} (\bibinfo{year}{2014}).
\newblock {\it \bibinfo{title}{Integrated {Resource} {Plan} ({IRP}) 2012}\/}.
\newblock \bibinfo{type}{Technical Report} Barbados Light and Power Co. Ltd.
  \bibinfo{address}{Bridgetown, Barbados}.
\bibitem[{{BPI}(2020)}]{bpi_port_2020}
\bibinfo{author}{{BPI}} (\bibinfo{year}{2020}).
\newblock \bibinfo{title}{Port {Statistics} 2010-2019}.
\bibitem[{Briguglio(1995)}]{briguglio_small_1995}
\bibinfo{author}{Briguglio, L.} (\bibinfo{year}{1995}).
\newblock \bibinfo{title}{Small island developing states and their economic
  vulnerabilities}.
\newblock {\it \bibinfo{journal}{World Development}\/},  {\it
  \bibinfo{volume}{23}\/}, \bibinfo{pages}{1615--1632}.
\bibitem[{Burnside-International(2014)}]{rj_burnside_international_preliminary_2014}
\bibinfo{author}{Burnside-International, R.} (\bibinfo{year}{2014}).
\newblock {\it \bibinfo{title}{Preliminary {Draft} {Waste} {Characterization}
  {Report}: {Mangrove} {Pond} {Green} {Energy} {Complex}}\/}.
\newblock \bibinfo{type}{Waste charaterization}
  \bibinfo{number}{300030295.0000} Barbados Sanitation Service Authority
  \bibinfo{address}{Bridgetown, Barbados}.
\bibitem[{Caricom(2018)}]{caricom_energy_2018}
\bibinfo{author}{Caricom, E.} (\bibinfo{year}{2018}).
\newblock {\it \bibinfo{title}{Energy {Report} {Card}}\/}.
\newblock \bibinfo{type}{Reports and {Studies}} The CARICOM Secretariat:
  Caribbean Centre for Renewable Energy and Energy Efficiency
  \bibinfo{address}{Georgetown, Guyanna}.
\bibitem[{Endresen(2003)}]{endresen_emission_2003}
\bibinfo{author}{Endresen, O.} (\bibinfo{year}{2003}).
\newblock \bibinfo{title}{Emission from international sea transportation and
  environmental impact}.
\newblock {\it \bibinfo{journal}{J. Geophys. Res.}\/},  {\it
  \bibinfo{volume}{108}\/}, \bibinfo{pages}{4560}.
\bibitem[{Espinasa et~al.(2016)Espinasa, Gischler, Humpert, Gonzalez~Torres \&
  Sucre}]{espinasa_achieving_2016}
\bibinfo{author}{Espinasa, R.}, \bibinfo{author}{Gischler, C.},
  \bibinfo{author}{Humpert, M.}, \bibinfo{author}{Gonzalez~Torres, C.}, \&
  \bibinfo{author}{Sucre, C.} (\bibinfo{year}{2016}).
\newblock \bibinfo{title}{Achieving {Sustainable} {Energy} in {Barbados}:
  {Energy} {Dossier}}, .
\bibitem[{Field et~al.(2015)Field, Rogers \& Sealy}]{field_effect_2015}
\bibinfo{author}{Field, D.}, \bibinfo{author}{Rogers, T.}, \&
  \bibinfo{author}{Sealy, A.} (\bibinfo{year}{2015}).
\newblock \bibinfo{title}{The effect of {Saharan} dust on {PV} system yields in
  the tropical environment of {Barbados}}.
\newblock In {\it \bibinfo{booktitle}{2015 {IEEE} 42nd {Photovoltaic}
  {Specialist} {Conference} ({PVSC})}\/} (pp. \bibinfo{pages}{1--4}).
\newblock \bibinfo{publisher}{IEEE}.
\bibitem[{Gay et~al.(2018)Gay, Rogers \& Shirley}]{gay_small_2018}
\bibinfo{author}{Gay, D.}, \bibinfo{author}{Rogers, T.}, \&
  \bibinfo{author}{Shirley, R.} (\bibinfo{year}{2018}).
\newblock \bibinfo{title}{Small island developing states and their suitability
  for electric vehicles and vehicle-to-grid services}.
\newblock {\it \bibinfo{journal}{Utilities Policy}\/},  {\it
  \bibinfo{volume}{55}\/}, \bibinfo{pages}{69--78}.
\bibitem[{Geidl et~al.(2006)Geidl, Favre-Perrod, Klöckl \&
  Koeppel}]{geidl_greenfield_2006}
\bibinfo{author}{Geidl, M.}, \bibinfo{author}{Favre-Perrod, P.},
  \bibinfo{author}{Klöckl, B.}, \& \bibinfo{author}{Koeppel, G.}
  (\bibinfo{year}{2006}).
\newblock \bibinfo{title}{A greenfield approach for future power systems}, .
\bibitem[{GOB(2019)}]{gob_barbados_2019}
\bibinfo{author}{GOB, G. o.~B.} (\bibinfo{year}{2019}).
\newblock {\it \bibinfo{title}{Barbados {National} {Energy} {Policy} 2019 -
  2030}\/}.
\newblock \bibinfo{type}{Offical policy document} \bibinfo{number}{BNEP} The
  Ministry of Energy \& Water Resources \bibinfo{address}{Trininty Business
  Complex - Bridgetown, Barbados}.
\bibitem[{Healey et~al.(2020)Healey, Beshilas, Coney \&
  Jackson}]{healey_energy_2020}
\bibinfo{author}{Healey, V.}, \bibinfo{author}{Beshilas, L.},
  \bibinfo{author}{Coney, K.}, \& \bibinfo{author}{Jackson, G.}
  (\bibinfo{year}{2020}).
\newblock {\it \bibinfo{title}{Energy {Snapshot} - {Barbados}}\/}.
\newblock \bibinfo{type}{Technical Report} \bibinfo{number}{NREL/FS-7A40-76636}
  National Renewable Energy Lab. (NREL), Golden, CO (United States).
\bibitem[{Henderson(2013)}]{henderson_sids_2013}
\bibinfo{author}{Henderson, V.} (\bibinfo{year}{2013}).
\newblock \bibinfo{title}{{SIDS} {DOCK}: {Facilitating} the {Transformation} of
  the {SIDS} {Energy} {Sector} {to} {Enable} {Climate} {Change} {Adaptation}:
  “25-50-25 {By} 2033”}.
\bibitem[{Henry et~al.(2015)Henry, Bridge, Henderson, {Kevin Keleher},
  Kirchhoff, {Geoff Goodwin}, Namugayi, Morris, Oaks, Dalrymple, Shrake, Ota,
  {Laura Azevedo}, Blue, Boucher, Boege, {Laura Hager}, {Tahja Mack},
  {Katherine Thompson} \& {Melissa Chavez}}]{henry_key_2015}
\bibinfo{author}{Henry, L.}, \bibinfo{author}{Bridge, J.},
  \bibinfo{author}{Henderson, M.}, \bibinfo{author}{{Kevin Keleher}},
  \bibinfo{author}{Kirchhoff, M.}, \bibinfo{author}{{Geoff Goodwin}},
  \bibinfo{author}{Namugayi, D.}, \bibinfo{author}{Morris, M.},
  \bibinfo{author}{Oaks, B.}, \bibinfo{author}{Dalrymple, O.},
  \bibinfo{author}{Shrake, S.}, \bibinfo{author}{Ota, A.},
  \bibinfo{author}{{Laura Azevedo}}, \bibinfo{author}{Blue, B.},
  \bibinfo{author}{Boucher, Z.}, \bibinfo{author}{Boege, S.},
  \bibinfo{author}{{Laura Hager}}, \bibinfo{author}{{Tahja Mack}},
  \bibinfo{author}{{Katherine Thompson}}, \& \bibinfo{author}{{Melissa Chavez}}
  (\bibinfo{year}{2015}).
\newblock {\it \bibinfo{title}{Key factors around ocean-based power in the
  {Caribbean} {Region}, via {Trinidad} and {Tobago}.}\/}
  volume~\bibinfo{volume}{50}.
\newblock \bibinfo{note}{Journal Abbreviation: Renewable and Sustainable Energy
  Reviews Publication Title: Renewable and Sustainable Energy Reviews}.
\bibitem[{Hilpert et~al.(2020)Hilpert, Dettner \&
  Al-Salaymeh}]{hilpert_analysis_2020}
\bibinfo{author}{Hilpert, S.}, \bibinfo{author}{Dettner, F.}, \&
  \bibinfo{author}{Al-Salaymeh, A.} (\bibinfo{year}{2020}).
\newblock \bibinfo{title}{Analysis of {Cost}-{Optimal} {Renewable} {Energy}
  {Expansion} for the {Near}-{Term} {Jordanian} {Electricity} {System}}.
\newblock {\it \bibinfo{journal}{Sustainability}\/},  {\it
  \bibinfo{volume}{12}\/}, \bibinfo{pages}{9339}.
\bibitem[{Hilpert et~al.(2021)Hilpert, Günther \&
  Söthe}]{hilpert_introducing_2021}
\bibinfo{author}{Hilpert, S.}, \bibinfo{author}{Günther, S.}, \&
  \bibinfo{author}{Söthe, M.} (\bibinfo{year}{2021}).
\newblock \bibinfo{title}{Introducing {Data} {Packages} for {Reproducible}
  {Workflows} in {Energy} {System} {Modeling}}.
\newblock {\it \bibinfo{journal}{Journal of Open Research Software}\/},  {\it
  \bibinfo{volume}{9}\/}, \bibinfo{pages}{6}.
\bibitem[{Hilpert \& Harewood(2021)}]{hilpert_oemof-barbados_2021}
\bibinfo{author}{Hilpert, S.}, \& \bibinfo{author}{Harewood, A.}
  (\bibinfo{year}{2021}).
\newblock \bibinfo{title}{oemof-barbados}.
\newblock \bibinfo{note}{Version 0.2 used for this work:
  \url{https://github.com/znes/oemof-barbados/releases/tag/v0.2}}.
\bibitem[{Hilpert et~al.(2018)Hilpert, Kaldemeyer, Krien, Günther, Wingenbach
  \& Plessmann}]{hilpert_open_2018}
\bibinfo{author}{Hilpert, S.}, \bibinfo{author}{Kaldemeyer, C.},
  \bibinfo{author}{Krien, U.}, \bibinfo{author}{Günther, S.},
  \bibinfo{author}{Wingenbach, C.}, \& \bibinfo{author}{Plessmann, G.}
  (\bibinfo{year}{2018}).
\newblock \bibinfo{title}{The {Open} {Energy} {Modelling} {Framework} (oemof) -
  {A} new approach to facilitate open science in energy system modelling}.
\newblock {\it \bibinfo{journal}{Energy Strategy Reviews}\/},  {\it
  \bibinfo{volume}{22}\/}, \bibinfo{pages}{16--25}.
\bibitem[{Hirth et~al.(2015)Hirth, Ueckerdt \&
  Edenhofer}]{hirth_integration_2015}
\bibinfo{author}{Hirth, L.}, \bibinfo{author}{Ueckerdt, F.}, \&
  \bibinfo{author}{Edenhofer, O.} (\bibinfo{year}{2015}).
\newblock \bibinfo{title}{Integration costs revisited – {An} economic
  framework for wind and solar variability}.
\newblock {\it \bibinfo{journal}{Renewable Energy}\/},  {\it
  \bibinfo{volume}{74}\/}, \bibinfo{pages}{925--939}.
\bibitem[{Hohmeyer(2015)}]{hohmeyer_100_2015}
\bibinfo{author}{Hohmeyer, O.} (\bibinfo{year}{2015}).
\newblock {\it \bibinfo{title}{A 100\% renewable {Barbados} and lower energy
  bills: {A} plan to change {Barbados}’ power supply to 100\% renewables and
  its possible benefits}\/}.
\newblock \bibinfo{type}{Discussion {Papers} 5} \bibinfo{number}{ISSN:
  2192-4597} Center for Sustainable Energy Systems (CSES/ZNES), System
  Integration Department \bibinfo{address}{Munketoft, Flensburg}.
\newblock \bibinfo{note}{Issue: ISSN: 2192-4597}.
\bibitem[{Hohmeyer(2017)}]{hohmeyer_economic_2017}
\bibinfo{author}{Hohmeyer, O.} (\bibinfo{year}{2017}).
\newblock {\it \bibinfo{title}{Economic {Analysis} to {Facilitate} the
  {Establishment} of a {Stable} {Price} for {Electricity} from {Renewable}
  {Sources}}\/}.
\newblock \bibinfo{type}{Technical Report} \bibinfo{number}{ME 36\_1\_2 T54}
  Global Sustainable Energy Consultants Ltd. \bibinfo{address}{Bridgetown,
  Barbados}.
\newblock \bibinfo{note}{Issue: ME 36\_1\_2 T54}.
\bibitem[{Hoyte(2016)}]{hoyte_shore--ship_2016}
\bibinfo{author}{Hoyte, D.} (\bibinfo{year}{2016}).
\newblock {\it \bibinfo{title}{Shore-to-ship {Power} for {Cruise} {Ships} on
  ‘100\% {Renewables}’ in the {Bridgetown} {Port}, {Barbados}.}\/}.
\newblock \bibinfo{type}{Master's {Thesis}} Centre for Resource Management and
  Environmental Studies (C.E.R.M.E.S) \bibinfo{address}{The University of the
  West Indies - Cave Hill Campus- Faculty of Science and Technology}.
\bibitem[{IADB(2010{\natexlab{a}})}]{iadb_support_2010}
\bibinfo{author}{IADB} (\bibinfo{year}{2010}{\natexlab{a}}).
\newblock {\it \bibinfo{title}{{Support} {for} {Sustainable} {Energy}
  {Framework} {for} {Barbados} ({SEFB}) {I}}\/}.
\newblock \bibinfo{type}{Technical Report} \bibinfo{number}{(BA-L1022)}
  Inter-American Developement Bank \bibinfo{address}{Barbados, Bridgetown}.
\bibitem[{IADB(2010{\natexlab{b}})}]{iadb_sustainable_2010}
\bibinfo{author}{IADB} (\bibinfo{year}{2010}{\natexlab{b}}).
\newblock {\it \bibinfo{title}{{Sustainable} {Energy} {Framework} {for}
  {Barbados}}\/}.
\newblock \bibinfo{type}{Plan of {Operation}} Inter-American Developement Bank
  \bibinfo{address}{Barbados, Bridgetown}.
\bibitem[{{IRENA}(2016)}]{irena_barbados_2016}
\bibinfo{author}{{IRENA}} (\bibinfo{year}{2016}).
\newblock {\it \bibinfo{title}{Barbados {Energy} {Roadmap}}\/}.
\newblock \bibinfo{type}{Technical Report} International Renewable Energy
  Agency \bibinfo{address}{United Arab Emirates}.
\bibitem[{IRENA(2019)}]{irena_renewable_2019}
\bibinfo{author}{IRENA} (\bibinfo{year}{2019}).
\newblock {\it \bibinfo{title}{Renewable {Power} {Generation} {Costs} in
  2018}\/}.
\newblock \bibinfo{type}{Technical Report} International Renewable Energy
  Agency \bibinfo{address}{Abu Dhabi}.
\bibitem[{Krien et~al.(2020)Krien, Schönfeldt, Launer, Hilpert, Kaldemeyer \&
  Pleßmann}]{krien_oemofsolphmodel_2020}
\bibinfo{author}{Krien, U.}, \bibinfo{author}{Schönfeldt, P.},
  \bibinfo{author}{Launer, J.}, \bibinfo{author}{Hilpert, S.},
  \bibinfo{author}{Kaldemeyer, C.}, \& \bibinfo{author}{Pleßmann, G.}
  (\bibinfo{year}{2020}).
\newblock \bibinfo{title}{oemof.solph—{A} model generator for linear and
  mixed-integer linear optimisation of energy systems}.
\newblock {\it \bibinfo{journal}{Software Impacts}\/},  {\it
  \bibinfo{volume}{6}\/}, \bibinfo{pages}{100028}.
\bibitem[{Lind et~al.(2018)Lind, Bjorn-Andersen, Watson, Ward, Bergmann,
  Rylander, Andersen, Hägg, Karlsson, Zerem, Haraldson, Pettersson, Lane,
  Carbajosa, Sancricca, Karlsson, Theodossiou \& Santén}]{lind_potential_2018}
\bibinfo{author}{Lind, M.}, \bibinfo{author}{Bjorn-Andersen, N.},
  \bibinfo{author}{Watson, R.}, \bibinfo{author}{Ward, R.},
  \bibinfo{author}{Bergmann, M.}, \bibinfo{author}{Rylander, R.},
  \bibinfo{author}{Andersen, T.}, \bibinfo{author}{Hägg, M.},
  \bibinfo{author}{Karlsson, M.}, \bibinfo{author}{Zerem, A.},
  \bibinfo{author}{Haraldson, S.}, \bibinfo{author}{Pettersson, S.},
  \bibinfo{author}{Lane, A.}, \bibinfo{author}{Carbajosa, J.},
  \bibinfo{author}{Sancricca, M.}, \bibinfo{author}{Karlsson, J.},
  \bibinfo{author}{Theodossiou, S.}, \& \bibinfo{author}{Santén, V.}
  (\bibinfo{year}{2018}).
\newblock {\it \bibinfo{title}{The {Potential} {Role} of {PortCDM} in {Cold}
  {Ironing}}\/}.
\bibitem[{{Macrotrends}(2021)}]{macrotrends_barbados_2021}
\bibinfo{author}{{Macrotrends}} (\bibinfo{year}{2021}).
\newblock \bibinfo{title}{Barbados {Greenhouse} {Gas} ({GHG}) {Emissions}
  1970-2021}.
\bibitem[{Marshall(2019)}]{marshall_bio-digestion_2019}
\bibinfo{author}{Marshall, R.} (\bibinfo{year}{2019}).
\newblock \bibinfo{title}{Bio-{Digestion}: {Benefits} to {Barbados}: {Central}
  {Bank} of {Barbados}'s {Biodigestion} {Conference} 2019: {PANEL}
  {DISCUSSION}}.
\bibitem[{Mitchell(2005)}]{mitchell_sugar_2005}
\bibinfo{author}{Mitchell, D.} (\bibinfo{year}{2005}).
\newblock {\it \bibinfo{title}{Sugar in the {Caribbean}: {Adjusting} to
  {Eroding} {Preferences}}\/}.
\newblock \bibinfo{type}{Working {Paper}} \bibinfo{number}{3902} The World
  Bank: Developement Prospects Group \bibinfo{address}{Washington, DC}.
\bibitem[{Mongird et~al.(2019)Mongird, Fotedar, Viswanathan, Koritarov,
  Balducci \& Hadjerioua}]{mongird_energy_2019}
\bibinfo{author}{Mongird, K.}, \bibinfo{author}{Fotedar, V.},
  \bibinfo{author}{Viswanathan, V.}, \bibinfo{author}{Koritarov, P.},
  \bibinfo{author}{Balducci, B.}, \& \bibinfo{author}{Hadjerioua, J.}
  (\bibinfo{year}{2019}).
\newblock {\it \bibinfo{title}{Energy {Storage} {Technology} and {Cost}
  {Characterization} {Report}}\/}.
\newblock \bibinfo{type}{Technical Report} US Department of Energy.
\bibitem[{{Mott MacDonald}(2021)}]{mott_macdonald_integrated_2021}
\bibinfo{author}{{Mott MacDonald}} (\bibinfo{year}{2021}).
\newblock {\it \bibinfo{title}{Integrated {Resource} \& {Resiliency} {Plan} for
  {Barbados}}\/}.
\newblock \bibinfo{type}{Technical Report} \bibinfo{number}{Draft Report}
  Inter-American Development Bank \bibinfo{address}{Brighton}.
\bibitem[{Niles \& Lloyd(2013)}]{niles_small_2013}
\bibinfo{author}{Niles, K.}, \& \bibinfo{author}{Lloyd, B.}
  (\bibinfo{year}{2013}).
\newblock \bibinfo{title}{Small {Island} {Developing} {States} ({SIDS}) \&
  energy aid: {Impacts} on the energy sector in the {Caribbean} and {Pacific}}.
\newblock {\it \bibinfo{journal}{Energy for Sustainable Development}\/},  {\it
  \bibinfo{volume}{17}\/}, \bibinfo{pages}{521--530}.
\bibitem[{OECD(2018)}]{oecd_making_2018}
\bibinfo{author}{OECD} (\bibinfo{year}{2018}).
\newblock {\it \bibinfo{title}{Making {Development} {Co}-operation {Work} for
  {Small} {Island} {Developing} {States}}\/}.
\newblock \bibinfo{type}{Technical Report} OECD Publishing
  \bibinfo{address}{Paris}.
\bibitem[{Pelling \& Uitto(2001)}]{pelling_small_2001}
\bibinfo{author}{Pelling, M.}, \& \bibinfo{author}{Uitto, J.~I.}
  (\bibinfo{year}{2001}).
\newblock \bibinfo{title}{Small island developing states: natural disaster
  vulnerability and global change}.
\newblock {\it \bibinfo{journal}{Global Environmental Change Part B:
  Environmental Hazards}\/},  {\it \bibinfo{volume}{3}\/},
  \bibinfo{pages}{49--62}.
\bibitem[{Pfenninger \& Staffell(2016)}]{pfenninger_long-term_2016}
\bibinfo{author}{Pfenninger, S.}, \& \bibinfo{author}{Staffell, I.}
  (\bibinfo{year}{2016}).
\newblock \bibinfo{title}{Long-term patterns of {European} {PV} output using 30
  years of validated hourly reanalysis and satellite data}.
\newblock {\it \bibinfo{journal}{Energy}\/},  {\it \bibinfo{volume}{114}\/},
  \bibinfo{pages}{1251--1265}.
\bibitem[{Rabl \& Spadaro(2000)}]{rabl_public_2000}
\bibinfo{author}{Rabl, A.}, \& \bibinfo{author}{Spadaro, J.~V.}
  (\bibinfo{year}{2000}).
\newblock \bibinfo{title}{Public health impacts of air pollution and
  implications for the energy system}.
\newblock {\it \bibinfo{journal}{Annual Review of Energy and the
  Environment}\/},  {\it \bibinfo{volume}{25}\/}, \bibinfo{pages}{601--627}.
\bibitem[{Robinson(2018)}]{robinson_adapting_2018}
\bibinfo{author}{Robinson, S.-A.} (\bibinfo{year}{2018}).
\newblock \bibinfo{title}{Adapting to climate change at the national level in
  {Caribbean} small island developing state}.
\newblock {\it \bibinfo{journal}{Island Studies Journal}\/},  {\it
  \bibinfo{volume}{13}\/}, \bibinfo{pages}{79--100}.
\bibitem[{Rogers(2017)}]{rogers_desktop_2017}
\bibinfo{author}{Rogers, T.} (\bibinfo{year}{2017}).
\newblock {\it \bibinfo{title}{A {Desktop} {Study} into the {Wind} {Resource}
  in {Barbados}}\/}.
\newblock \bibinfo{type}{Technical Report} The University of the West Indies
  Cave Hill Campus \bibinfo{address}{Bridgetown, Barbados}.
\bibitem[{Schuhmann et~al.(2019)Schuhmann, Skeete, Waite, Lorde,
  Bangwayo-Skeete, Oxenford, Gill, Moore \& Spencer}]{schuhmann_visitors_2019}
\bibinfo{author}{Schuhmann, P.~W.}, \bibinfo{author}{Skeete, R.},
  \bibinfo{author}{Waite, R.}, \bibinfo{author}{Lorde, T.},
  \bibinfo{author}{Bangwayo-Skeete, P.}, \bibinfo{author}{Oxenford, H.~A.},
  \bibinfo{author}{Gill, D.}, \bibinfo{author}{Moore, W.}, \&
  \bibinfo{author}{Spencer, F.} (\bibinfo{year}{2019}).
\newblock \bibinfo{title}{Visitors’ willingness to pay marine conservation
  fees in {Barbados}}.
\newblock {\it \bibinfo{journal}{Tourism Management}\/},  {\it
  \bibinfo{volume}{71}\/}, \bibinfo{pages}{315--326}.
\bibitem[{Simmons \& Associates(2015)}]{simmons_and_associates_barabados_2015}
\bibinfo{author}{Simmons}, \& \bibinfo{author}{Associates}
  (\bibinfo{year}{2015}).
\newblock {\it \bibinfo{title}{Barabados {Waste} {Characterization} {Study} for
  {Barbados}}\/}.
\newblock \bibinfo{type}{Final {Report}} Project management coordination unit:
  Ministry of Enironment and Drainage \bibinfo{address}{Bridgetown, Barbados}.
\bibitem[{Staffell \& Pfenninger(2016)}]{staffell_using_2016}
\bibinfo{author}{Staffell, I.}, \& \bibinfo{author}{Pfenninger, S.}
  (\bibinfo{year}{2016}).
\newblock \bibinfo{title}{Using bias-corrected reanalysis to simulate current
  and future wind power output}.
\newblock {\it \bibinfo{journal}{Energy}\/},  {\it \bibinfo{volume}{114}\/},
  \bibinfo{pages}{1224--1239}.
\bibitem[{Tang et~al.(2020)Tang, Ramacher, Moldanová, Matthias, Karl,
  Johansson, Jalkanen, Yaramenka, Aulinger \& Gustafsson}]{tang_impact_2020}
\bibinfo{author}{Tang, L.}, \bibinfo{author}{Ramacher, M. O.~P.},
  \bibinfo{author}{Moldanová, J.}, \bibinfo{author}{Matthias, V.},
  \bibinfo{author}{Karl, M.}, \bibinfo{author}{Johansson, L.},
  \bibinfo{author}{Jalkanen, J.-P.}, \bibinfo{author}{Yaramenka, K.},
  \bibinfo{author}{Aulinger, A.}, \& \bibinfo{author}{Gustafsson, M.}
  (\bibinfo{year}{2020}).
\newblock \bibinfo{title}{The impact of ship emissions on air quality and human
  health in the {Gothenburg} area – {Part} 1: 2012 emissions}.
\newblock {\it \bibinfo{journal}{Atmos. Chem. Phys.}\/},  {\it
  \bibinfo{volume}{20}\/}, \bibinfo{pages}{7509--7530}.
\bibitem[{{UNDP}(2009)}]{undp_preliminary_2009}
\bibinfo{author}{{UNDP}} (\bibinfo{year}{2009}).
\newblock {\it \bibinfo{title}{Preliminary {Assessment} of {Bioenergy}
  {Production} in the {Caribbean}}\/}.
\newblock \bibinfo{type}{Technical Report} United Nations Developement
  Programme \bibinfo{address}{Barbados and the OECS}.
\bibitem[{UNDP(2018)}]{undp_high-level_2018}
\bibinfo{author}{UNDP} (\bibinfo{year}{2018}).
\newblock {\it \bibinfo{title}{High-level {Political} {Forum} on {Sustainable}
  {Developement}}\/}.
\newblock \bibinfo{type}{Technical Report} United Nations Developement
  Programme.
\bibitem[{UNDP(2020)}]{undp_barbados_2020}
\bibinfo{author}{UNDP} (\bibinfo{year}{2020}).
\newblock \bibinfo{title}{Barbados {\textbar} {UNDP} {Climate} {Change}
  {Adaptation}}.
\newblock \bibinfo{note}{Library Catalog: www.adaptation-undp.org}.
\bibitem[{Viana et~al.(2020)Viana, Rizza, Tobías, Carr, Corbett, Sofiev,
  Karanasiou, Buonanno \& Fann}]{viana_estimated_2020}
\bibinfo{author}{Viana, M.}, \bibinfo{author}{Rizza, V.},
  \bibinfo{author}{Tobías, A.}, \bibinfo{author}{Carr, E.},
  \bibinfo{author}{Corbett, J.}, \bibinfo{author}{Sofiev, M.},
  \bibinfo{author}{Karanasiou, A.}, \bibinfo{author}{Buonanno, G.}, \&
  \bibinfo{author}{Fann, N.} (\bibinfo{year}{2020}).
\newblock \bibinfo{title}{Estimated health impacts from maritime transport in
  the {Mediterranean} region and benefits from the use of cleaner fuels}.
\newblock {\it \bibinfo{journal}{Environment International}\/},  {\it
  \bibinfo{volume}{138}\/}, \bibinfo{pages}{105670}.
\bibitem[{Zis et~al.(2014)Zis, North, Angeloudis, Ochieng \&
  Bell}]{zis_evaluation_2014}
\bibinfo{author}{Zis, T.}, \bibinfo{author}{North, R.},
  \bibinfo{author}{Angeloudis, P.}, \bibinfo{author}{Ochieng, W.}, \&
  \bibinfo{author}{Bell, M.} (\bibinfo{year}{2014}).
\newblock \bibinfo{title}{Evaluation of cold ironing and speed reduction
  policies to reduce ship emissions near and at ports}.
\newblock {\it \bibinfo{journal}{Maritime Economics \& Logistics}\/},  {\it
  \bibinfo{volume}{16}\/}, \bibinfo{pages}{371--398}.

\end{thebibliography}

\end{document}